\documentclass[11pt]{article}
\usepackage[english,ruled,vlined]{algorithm2e}
\SetKwInput{KwInput}{Inputs}
\SetKwInput{KwReturn}{Return}
\SetKwInput{KwRequire}{Require}
\usepackage[noend]{algpseudocode}

\usepackage{pgfplots}
\pgfplotsset{compat=newest}
\usetikzlibrary{shapes}
\usetikzlibrary{plotmarks}
\usepackage{footmisc}
\usepackage{multirow}
\usepackage{listings}
\usepackage{booktabs}

\usepackage{amssymb}
\usepackage{amsfonts}
\usepackage{amsmath}
\usepackage{textcomp}
\usepackage{amsthm}
\usepackage{mathbbol}
\usepackage{stmaryrd}
\usepackage{mathrsfs}
\usepackage{graphicx}
\usepackage{subfigure}
\usepackage{amsbsy}
\usepackage{xcolor}
\usepackage{hyperref}
\usepackage{tikz}
\usetikzlibrary{arrows,shapes}
\usetikzlibrary{shapes.geometric}
\usetikzlibrary{shapes.arrows}

\usepackage[normalem]{ulem}
\usepackage{float}
\usepackage{dsfont}
\usepackage{empheq}
\usepackage{cleveref}
\usepackage[toc,title,page]{appendix}

\definecolor{mygreen}{rgb}{0,0.7,0}

\definecolor{mygreen}{rgb}{0,0.4,0}

\newcommand{\GMcorr}%{\textcolor{red}}

\title{\sf  Micromagnetic simulations of the size dependence of the Curie temperature in ferromagnetic nanowires and nanolayers}
\author{Cl\'ementine Court\`es$^{1,2}$ \and Matthieu Boileau$^{1,2}$ \and Rapha\"el C\^ote$^{1,3}$\and Paul-Antoine Hervieux$^4$\and Giovanni Manfredi$^4$}

\begin{document}
\maketitle

\footnotetext[1]{University of Strasbourg,
CNRS, IRMA UMR 7501, F-67000
Strasbourg, France}
\footnotetext[2]{INRIA Nancy Grand-Est, MACARON Team, France}
\footnotetext[3]{University of Strasbourg, Institute for Advanced Study, F-67000 Strasbourg, France}
\footnotetext[4]{University of Strasbourg, CNRS, IPCMS UMR 7504, F-67000 Strasbourg, France}

\begin{abstract}
We solve the Landau-Lifshitz-Gilbert equation in the finite-temperature regime, where thermal fluctuations are modeled by a random magnetic field whose variance is proportional to the temperature.
By rescaling the temperature proportionally to the computational cell size $\Delta x$ ($T \to T\,\Delta x/a_{\text{eff}}$, where $a_{\text{eff}}$ is the lattice constant) [M. B. Hahn, J. Phys. Comm., 3:075009, 2019], we obtain Curie temperatures $T_{\text{C}}$ that are in line with the experimental values for cobalt, iron and nickel.
For finite-sized objects such as nanowires (1D) and nanolayers (2D), the Curie temperature varies with the smallest size $d$ of the system. We show that the difference between the computed finite-size $T_{\text{C}}$ and the bulk $T_{\text{C}}$ follows a power-law of the type: $(\xi_0/d)^\lambda$, where $\xi_0$ is the correlation length at zero temperature, and $\lambda$ is a critical exponent. We obtain values of $\xi_0$ in the nanometer range, also in accordance with other simulations and experiments. The computed critical exponent is close to $\lambda=2$ for all considered materials and geometries. This is the expected result for a mean-field approach, but slightly larger than the values observed experimentally.
\end{abstract}

\tableofcontents

%\bs
\quad


%\noindent {\bf 2020 AMS subject classifications: }35Q60, 78M20 \textcolor{red}{+ classification physique}

%%%%%%%%%%%%%%%%%%%%%%%%%%%%%%%%%%%%%%%%%%%
\section{Introduction}

Interest in nano-scale ferromagnetic objects (nanowires, nanolayers, etc.) has grown dramatically in recent years, as these objects are integrated into many current devices with the aim of storing, reading and writing digital information. 
Hence, there is a growing need for accurate numerical simulations capable of predicting the behaviour of such objects, and possibly predict new properties and features.
Computational models of ferromagnetic materials can be roughly divided into two families. On the one hand, atomistic models such as \verb"VAMPIRE" \cite{VAMPIRE} describe the magnetic interactions microscopically, at the natural atomic length scale of the material. Although in principle very accurate, they demand a considerable computational cost and are thus limited to relatively small systems.

In contrast, micromagnetism describes the structure and dynamics of a ferromagnetic object at an intermediate mesoscopic scale, averaging over a large number of atoms. Micromagnetic codes, based on the Landau-Lifshitz-Gilbert (LLG) equation, describe the dynamics of the  average magnetic moment $\boldsymbol{m}(t, \boldsymbol{x})$, which is a continuous function of the spatial coordinate $\boldsymbol{x}$. Notable projects in computational micromagnetism are the \verb"OOMMF" project \cite{OOMMF} (Object Oriented MicroMagnetic Framework) for the development of a public micromagnetic program in C++, the \verb"mumax3" project \cite{mumax3}, a GPU-accelerated micromagnetic simulation program, or else the \verb"tetmag" project, a 3D micromagnetic finite-element simulation software \cite{hertel_tetmag_2023}.

Here, we will adopt the micromagnetic approach (LLG equation) to study the influence of the temperature on some fundamental properties of both 1D (nanowires) and 2D (nanolayers) ferromagnetic nano-objects, for which thermal effects may become important. 
In order to model thermal fluctuations in the context of micromagnetics, in 1963
Brown \cite{Brown_1963} proposed to add a stochastic term to the LLG equation, in the form of a randomly fluctuating magnetic field with zero mean and variance that is proportional to the temperature.

However, the Brown method suffers from a fundamental problem: while the LLG equation is valid at a mesoscopic scale and the corresponding magnetic moment $\boldsymbol{m}(t, \boldsymbol{x})$ represents an average over many atomic spins, thermal fluctuations occur at the atomic level. Hence, one is mixing two different levels of descriptions in the same LLG equation: the mesoscopic level for the deterministic terms, and the microscopic level for the stochastic terms. Then, if applied without further corrections, this procedure entails that temperature effects (for instance, the numerically-calculated Curie temperature) depend on the computational cell, which is obviously a spurious result.

Indeed, the computational cell size $\Delta x$ is noticeably larger than the physical lattice constant $a_{\rm eff}$. Thermal fluctuations occur at the length scale $a_{\rm eff}$, but are necessarily implemented at the scale $\Delta x$ in the micromagnetic codes. 
This induces an error in the computed properties, in particular near the Curie temperature $T_{\text{C}}$, which can be overestimated by one order of magnitude or more \cite{Grinstein_Koch_2003}.

In order to mitigate this spurious effect, several strategies have been proposed. 
In one approach \cite{Kirschner_2005,Kirschner2006}, the magnetization at saturation $M_s$ is scaled with the temperature and the computational cell size, using a Bloch-like law that is similar to the well-known temperature dependence of $M_s$. By comparing the micromagnetic results to those obtained with an atomistic model, the authors of \cite{Kirschner_2005,Kirschner2006} obtain a difference of less than $1\%$ in the estimation of the equilibrium magnetization, for a temperature $T=0.38\, T_{\text{C}}$ and a computational cell size $\Delta x = 1.5\; \rm nm$.

Another approach \cite{Grinstein_Koch_2003} consists in defining a rescaled temperature to take into account the fact that thermal averages (coarse graining) are performed on a computational cell that is larger than the lattice constant. In particular, the effective exchange constant varies with the size of the coarse-graining block, and this dependence should be taken into account. Grinstein and Koch \cite{Grinstein_Koch_2003} used renormalization-group techniques to unravel this dependence. In the same spirit, Hahn \cite{Hahn_2019} proposed a simple scaling law between the physical temperature and a "numerical" temperature to be used in the micromagnetic code, which depends on the ratio between the computational cell size and the lattice constant. This method was tested for nickel, cobalt and iron objects using the \verb"OOMMF" code \cite{OOMMF}, and yielded Curie temperature that were virtually independent on the computational cell and very close to the experimental values.

A possible limitation of the LLG approach is that the amplitude of the local magnetic moment $|\boldsymbol{m}(t, \boldsymbol{x})|$ remains constant in time, which is not necessarily true at high temperatures, notably near $T_{\text{C}}$. Chubykalo-Fesenko et al. \cite{Chubykalo2006} have investigated this issue using an atomistic time-dependent model and indeed they found that the modulus of the magnetization varies in time (see figure 1 in \cite{Chubykalo2006}). However, this variation is limited to a dip during an initial transient, after which $|\boldsymbol{m}(t, \boldsymbol{x})|$ recovers approximately its initial value. As our results are obtained by taking time-averages at longer times, this variation should not be too significant. But indeed, when studying transient phenomena, it may be necessary to take this effect into account, for instance by using a Landau-Lifshitz-Bloch approach, as suggested in \cite{Chubykalo2006,Atxitia2017}.

In the present work, we adopt Hahn's method to model thermal effects \cite{Hahn_2019} and use it to study the dependence of the magnetization law (in particular the Curie temperature) with the size of the system  under consideration. We will focus on two nano-objects, namely one-dimensional (1D) nanowires and 2D nanolayers. 
Theoretical considerations \cite{Fisher1972} indicate that the Curie temperature follows a power-law of the type:
\[
\frac{T_{\text{C}}(\infty)-T_{\text{C}}(d)}{T_{\text{C}}(\infty)}=\left(\frac{\xi_0}{d}\right)^{\lambda},
\]
where $d$ is the smallest size of the system, $T_{\text{C}}(\infty)$ and $T_{\text{C}}(d)$ are the Curie temperatures of the bulk and of the finite system respectively, $\xi_0$ is the correlation length at zero temperature, and $\lambda$ is the critical exponent.
The main purpose of this work will be to validate the above law and obtain the exponent and the correlation length from micromagnetic simulations with thermal effects.

From an experimental point of view, several works considered this problem. 
Early studies on thin nickel films, cobalt films and $\rm Co_1 Ni_9$ alloys yielded critical exponents $\lambda=1.25$, 1.34 and 1.39, respectively \cite{Schneider1990,Huang1993}, while measurements on nickel films \cite{Li1992} revealed an exponent $\lambda=1.4$.
Later work on nickel nanowires \cite{Sun_2000}, with diameters ranging from $30\, \rm nm$ to $500\, \rm nm$,  yielded  $\xi_0=2.2 \; \rm nm$ and $\lambda=0.94$.
More recent works \cite{Zhang2001,Wang2011} mention larger values for the exponent, up to $\lambda=2.8$. 

On the theory front, statistical estimations based on a first-principles-based Monte Carlo approach yielded a critical exponent $\lambda=1.47$ for $\rm Pb(Zr_{0.5}\,Ti_{0.5}\, O_3)$ (PZT) thin films \cite{Almahmoud2010}. Similar values were obtained using other approaches \cite{Yang2005}.
These exponents should be compared to the theoretical values predicted by the 3D Heisenberg and Ising models (respectively, $\lambda=1.4$ and $\lambda=1.58$).
\GMcorr{Atomistic calculations \cite{Hovorka2012} of FePt nanoparticles (cylindrically shaped, with height and diameter in the range $2-9\, \rm nm$), performed using either long-range or nearest-neighbor exchange, showed values of the critical exponent in the range $\lambda=1.18-1.25$.
}
In contrast, another atomistic mean-field model \cite{Penny2019} yielded larger values, close to $\lambda = 2$ (range: $\lambda \in [1.82-2.17]$), for magnetite nanoparticles of different shapes and sizes. 
As we shall see, our own work suggests a critical exponent close to $\lambda \approx 2$, for both nanowires (1D) and nanolayers (2D). The observed correlation length is found to be $\xi_0 \approx 3$~nm for nanowires and $\xi_0 \approx 1.6$~nm for nanolayers. 

We further note that our results are obtained using a time-dependent model, in contrast to statistical and Monte Carlo approaches used in other studies \cite{Almahmoud2010}. In other words, we solve the time-dependent LLG equation with thermal effects and, once a fluctuating equilibrium is reached, we measure the relevant magnetic properties by performing ensemble averages over many realizations and/or time averages over a certain duration. 
This approach is less computationally expensive than fully atomistic simulations.
In addition, it is amenable to investigating the {\em dynamical} properties of magnetism, such as the propagation of domain walls and other transient effects, which will make the object of future work. Here, we have used this time-dependence to show that statistical fluctuations explode near $T_{\text{C}}$, confirming the presence of a phase transition at that temperature.

The present paper is organized as follows. Section~\ref{sec:math_eq} details the mathematical setting, namely the LLG equation at finite temperature. After recalling the various terms involved in the effective magnetic field in Sec.~\ref{subsec:H_eff}, the following subsections are devoted to the modelling of thermal effects through a stochastic magnetic field (Sec.~\ref{subsection_Thermal_fluctuations}) and to the implementation of the temperature scaling \cite{Hahn_2019} (Sec.~\ref{sec:Tscaling}). Section \ref{sec:numeric_scaling} contains the details of the numerical scheme.
Section~\ref{sec:validation_code} is devoted to the validation of the numerical code with several test cases (Sec.~\ref{sec:testdetails}), clearly proving that the Curie temperature does not depend on the computational cell size or the time step (Sec.~\ref{sec:independence}). The validity of the Bloch law (at low temperatures, $T \ll T_{\text{C}}$) and Curie law (at $T \lesssim T_{\text{C}}$) are also tested (Sec.~\ref{sec:bloch}).
Finally, Section~\ref{sec:size_effects} contains the main results of this work, namely the size dependence of the Curie temperature for two types of nano-objects: 1D nanowires and 2D nanolayers. Conclusions are drawn in Section~\ref{Conclusion}.

\section{Micromagnetic model at finite temperature}\label{sec:math_eq}

The present numerical study focuses on a ferromagnetic domain modeled either as an infinite nanowire along the $\boldsymbol{e}_x$ axis, where $(\boldsymbol{e}_x, \boldsymbol{e}_y, \boldsymbol{e}_z)$ is the canonical orthonormal basis in $\mathbb{R}^3$, or an infinite nanolayer in the $(\boldsymbol{e}_x, \boldsymbol{e}_y)$ plane. For all times $t\geq 0$ and  positions $\boldsymbol{x}\in \mathbb{R}^3$, let $\boldsymbol{m}(t,\boldsymbol{x})\in \mathbb{S}^2=\{\boldsymbol{m}\in\mathbb{R}^3, \|\boldsymbol{m}\|=1\}$ be the magnetic moment vector field normalized to the saturation magnetization $M_s$. Here, $\mathbb{S}^2$ is the unit sphere. The precession dynamics of  $\boldsymbol{m}(t,\boldsymbol{x})$ is described by the  Landau-Lifshitz-Gilbert (LLG) equation:
\begin{equation}\label{LLGeq}
\frac{\partial\boldsymbol{m}}{\partial t}=-\gamma_0\,\boldsymbol{m}\times\boldsymbol{H}_{\text{eff}}-\gamma_0\alpha\,\boldsymbol{m}\times\left(\boldsymbol{m}\times\boldsymbol{H}_{\text{eff}}\right),\\
\end{equation}
where $\boldsymbol{H}_{\text{eff}}$ is the effective magnetic field. Here, $\gamma_0=\gamma \mu_0 >0$ is the scaled gyromagnetic ratio, with $\gamma = e/2m$ (where $e>0$ and $m$ are the charge and mass of the electron, respectively), and $\alpha>0$ is the dimensionless damping constant \cite{Gilbert2004}. Table \ref{table_constantes_physiques} lists all the physical variables used in this work, their units, and their numerical values.

\subsection{Effective magnetic field $\boldsymbol{H}_{\text{eff}}$}\label{subsec:H_eff}

The effective magnetic field $\boldsymbol{H}_{\text{eff}}$ results from the sum of  the exchange field $\boldsymbol{H}_{\text{exch}}$ and the anisotropy field $\boldsymbol{H}_{\text{ani}}$:
\[
\boldsymbol{H}_{\text{eff}}= \boldsymbol{H}_{\text{exch}}+\boldsymbol{H}_{\text{ani}}\;.
\]
The exchange field is due to the Heisenberg exchange interaction and is written as 
\begin{equation}
\boldsymbol{H}_{\text{exch}} = \frac{2A}{\mu_0 M_s} \Delta \boldsymbol{m},
\end{equation}
with $A>0$ the exchange constant and $\mu_0>0$ the vacuum permeability (see Table \ref{table_constantes_physiques}).  

The anisotropy field is due to the existence of preferred directions in the magneto-crystalline  structure of the material. Throughout the following, two cases of anisotropy field will be considered: uniaxial anisotropy (for cobalt systems) and cubic anisotropy (for nickel and iron systems), whose expressions are given below:

\begin{subequations}
\begin{equation}
\boldsymbol{H}_{\text{ani, uniaxial}}=\frac{2K}{\mu_0M_s}(\boldsymbol{e}_x\cdot\boldsymbol{m})\boldsymbol{e}_x,\label{uniaxial}
\end{equation}
\begin{equation}
\boldsymbol{H}_{\text{ani, cubic}}=-\frac{2K}{\mu_0M_s}\sum_{(i,j,k)\in I}\left((\boldsymbol{e}_j\cdot\boldsymbol{m})^2+(\boldsymbol{e}_k\cdot\boldsymbol{m})^2+(\boldsymbol{e}_j\cdot\boldsymbol{m})^2(\boldsymbol{e}_k\cdot\boldsymbol{m})^2\right)(\boldsymbol{e}_i\cdot\boldsymbol{m})\boldsymbol{e}_i,\label{cubic}
\end{equation}
\end{subequations}
where $I=\{(x, y, z), (y, z, x), (z, x, y)\}$ and $K>0$ is the anisotropy constant (assumed identical in all three directions of space for the cubic case), see Table~\ref{table_constantes_physiques}.

In the following, we will assume that the ferromagnetic domain is not subjected to any external magnetic field, so that no Zeeman energy is present.  The demagnetizing field (due to the magnetic field generated by the nanowire or nanolayer itself) and the dipolar interactions are also not taken into account.

\begin{table}
\begin{center}
\begin{tabular}{p{2cm}p{6cm}p{6cm}}
\toprule
\multicolumn{3}{c}{\textbf{Universal constants}}\\
\hline
$\gamma$& gyromagnetic ratio&$1.76\times10^{11}$ rad s$^{-1}$T$^{-1}$\\
$\mu_0$&vacuum permeability&$4\pi\times10^{-7}$ NA$^{-2}$\\
$\gamma_0$&rescaled gyromagnetic ratio&$\gamma\mu_0$ mA$^{-1}$s$^{-1}$\\
$k_B$&Boltzmann constant&$1.38\times10^{-23}$ JK$^{-1}$\\
&&\\
\end{tabular}
\begin{tabular}{lllll}
\toprule
\multicolumn{2}{c}{\textbf{Magnetic bulk parameters}}&Cobalt&Iron&Nickel\\
\hline
$A$&exchange constant&$3 \times 10^{-11}$ Jm$^{-1}$&$2.1 \times 10^{-11}$ Jm$^{-1}$&$9 \times 10^{-12}$ Jm$^{-1}$\\
$K$&anisotropy constant&$5.2\times10^{5}$ Jm$^{-3}$&$4.8\times10^{4}$ Jm$^{-3}$&$-5.7\times10^{3}$ J.m$^{-3}$\\
$M_s$&saturation magnetization&$1.4\times 10^{6}$ Am$^{-1}$&$1.7\times 10^{6}$ Am$^{-1}$&$4.9\times 10^{5}$ Am$^{-1}$\\
$a_{\text{eff}}$&lattice constant&$0.25$ nm&$0.286$ nm&$0.345$ nm\\
$\boldsymbol{H}_{\text{ani}}$&anisotropy field&uniaxial&cubic&cubic\\
$T_{\text{C}}$ & Curie temperature & $1388$ K& $1043$ K & $627$ K\\
$\alpha$ & Damping parameter & 0.5 & 0.5 & 0.5 \\
&&&&\\
\bottomrule
\end{tabular}
\caption{Top: Values of the universal constants used in this work. Bottom: Magnetic parameters for bulk cobalt, iron, and nickel. Sources: \cite{Kittelbook2018,Davey1925,Ono1988}. }
\label{table_constantes_physiques}
\end{center}
\end{table}

\subsection{Thermal fluctuations} \label{subsection_Thermal_fluctuations}

The deterministic LLG equation considered above is valid in the zero-temperature regime. However, thermal effects obviously influence the magnetic properties, first and foremost by cancelling out the spontaneous magnetization of a ferromagnetic material above a certain critical temperature (Curie temperature $T_{\text{C}}$). The material then goes from a ferromagnetic to a paramagnetic state when $T_{\text{C}}$ is crossed. 

In order to model thermal fluctuations, an additional random field is added to the effective magnetic field, following the idea of W.F. Brown \cite{Brown_1963}, so that we have:
\[
\boldsymbol{H}_{\text{eff}}= \boldsymbol{H}_{\text{exch}}+\boldsymbol{H}_{\text{ani}} + \boldsymbol{H}_{\text{stocha}} \;.
\]
The random thermal field  $\boldsymbol{H}_{\text{stocha}}$ is an isotropic Gaussian white-noise vector process of variance $\nu^2\in\mathbb{R}$. More precisely,  $\boldsymbol{H}_{\text{stocha}}$ may be written as: 
$\boldsymbol{H}_{\text{stocha}}(t)dt= \nu \,d\boldsymbol{W}(t)$, where $\boldsymbol{W}(t)= \big(W_1 (t), W_2 (t), W_3 (t)\big)^\text{T}$ is a classical time-continuous Wiener process. 

\paragraph{Wiener process.} The main properties of the stochastic field $\boldsymbol{W}(t)$ (and so $\boldsymbol{H}_{\text{stocha}}(t)$) are listed below, denoting $\langle \cdot \rangle=\mathbb{E}(\cdot)$ the statistical average \cite{Klughertz_article_2014, dAquino_et_al, Ragusa_et_al}:

\begin{itemize}
\item
Space homogeneity: $\boldsymbol{W}$ only depends on $t$ and not on $\boldsymbol{x}$,
\item
Continuous time random process: $\boldsymbol{W}(t), \forall t\geq0$,
\item
Null mean: $\langle \boldsymbol{W}(t)\rangle = 0, \forall t\geq0$,
\item
Decorrelated spatial components and vanishingly small autocorrelation time:  $\langle W_i(t)W_j(t')\rangle = \delta_{ij}\delta(t-t')$, $\forall i,j \in\{1, 2, 3\}$ (the indices of spatial components) and $t, t'\in\mathbb{R}^+$. Here, $\delta(\cdot)$ is the Dirac distribution and $\delta_{ij}$ is the Kronecker symbol.
\end{itemize}

\paragraph{Variance.} The standard deviation $\nu$ (and thus the variance $\nu^2$) is directly related to the temperature $T$ thanks to the following relation, obtained by the fluctuation-dissipation theorem \cite{Brown_1963, Ragusa_et_al} adapted here to the expression of Eq.~\eqref{LLGeq}:

\begin{equation}
\nu^2 = \frac{2\alpha k_B T}{\gamma_0\mu_0 M_s V},
\label{eq:variance}
\end{equation}
where $k_B$ is the Boltzmann constant (see Table~\ref{table_constantes_physiques}) and $V$ stands for a characteristic volume that depends on the internal crystalline structure of the material (which can be a Face-Centered Cubic lattice (FCC) as in the case of nickel, a Body Centered Cubic lattice (BCC) as in iron, or a Hexagonal Close-Packed lattice (HCP) for cobalt). Following the notation of \cite{Hahn_2019}, the shorter lattice distance is called the lattice constant $a_{\text{eff}}$ and the corresponding characteristic volume is $V=a_{\text{eff}}^3$.

\paragraph{Stochastic Landau-Lifshitz-Gilbert equation.} Consequently, the Landau-Lifshitz-Gilbert equation \eqref{LLGeq}  is modified to take the form of a stochastic partial differential equation (SPDE) 
\begin{equation}
d\boldsymbol{m}=-\gamma_0\boldsymbol{m}\times\left(\boldsymbol{H}_{\text{eff}}dt+\nu d\boldsymbol{W}\right)-\gamma_0\alpha\boldsymbol{m}\times\left[\boldsymbol{m}\times\left(\boldsymbol{H}_{\text{eff}}dt+\nu d\boldsymbol{W}\right)\right].
\label{LL_stocha}
%\tag{LL$_{\text{stocha}}$}
\end{equation}
All physical constants used in the forthcoming simulations are summarized in Table~\ref{table_constantes_physiques}.

In order to preserve the constraint that the magnetic moment $\boldsymbol{m}$ be on the unit sphere, i.e. $\|\boldsymbol{m}(t,\boldsymbol{x})\|=1$, $\forall \boldsymbol{x}\in\mathbb{R}^3$ and $t\geq 0$, we interpret the above stochastic LLG equation in the Stratonovich sense; see \cite{Labbe_Lelong} for the issues posed by using It\^o's approach.

\subsection{Temperature scaling with the computational cell size} \label{sec:Tscaling}
The variance of the stochastic magnetic field  used to model thermal effects depends not only on the temperature $T$, but also on the characteristic volume $V$, see Eq. (\ref{eq:variance}). In principle, this volume is related to the lattice constant, i.e. $V= a_{\text{eff}}^3$, but in practice $a_{\text{eff}}$ is much smaller than the spatial step $\Delta x$ used in the simulations. However, if one takes instead $V=\Delta x^3$, the simulation results will depend on the computational cell size, which is not an acceptable situation. At a fundamental level, this is due to the fact that the effective exchange constant varies with the size of the coarse-graining block \cite{Grinstein_Koch_2003}.

In order to suppress this unwanted numerical effect, we follow the procedure recently suggested by Hahn 
\cite{Hahn_2019}, which relies on a scaling of the temperature with the spatial step  $\Delta x$. The argument goes as follows: near the Curie temperature, the ferromagnetic behaviour disappears because the energy density of the thermal fluctuations $k_B T/V$, which favor random orientations of the spin, becomes similar to the energy density of the exchange interactions $A |\nabla \boldsymbol{m}|^2$, which favor magnetic order.  Hence, we write:
\begin{equation}
\frac{k_B T}{a^3} \sim \frac{A |\boldsymbol{m}|^2}{a^2},
\label{eq:Tscaling}
\end{equation}
where $a$ is a characteristic length that can be either the lattice constant $a_{\text{eff}}$ or the computational cell size $\Delta x$. From Eq. (\ref{eq:Tscaling}) it is clear that, in order for the magnetic moment to be independent on the averaging volume $a^3$, the temperature must scale as $T \sim a$.

Therefore, we define a "numerical" temperature as
\[
T_{\text{num}} = \frac{\Delta x}{a_{\text{eff}}} \,T.
\]
According to the above considerations, taking a volume $V=\Delta x^3$ together with the temperature $T_{\text{num}}$ should give results that are independent on $\Delta x$ and identical to those obtained using the ``physical'' volume $V= a_{\text{eff}}^3$ and the real temperature $T$. This approach was recently tested by Hahn \cite{Hahn_2019}, who indeed observed near independence of $T_{\text{C}}$ on the cell size up to $\Delta x=4 \; \rm nm$ for ferromagnetic thin films.
Following this procedure, the numerical variance of the fluctuating field is defined as
\begin{equation}
\nu_{\text{num}}^2 =\frac{2\alpha k_B\frac{\Delta x}{a_{\text{eff}}}T}{\gamma_0\mu_0M_s\Delta x^3},
\label{eq_variance_num}
\end{equation}
which replaces  Eq. (\ref{eq:variance}) in the simulations.
This expression may also be interpreted as meaning that the numerical characteristic volume to be taken into account is $V=a_{\text{eff}}\Delta x^2$.

\section{Computational method}\label{sec:numeric_scaling}

\begin{figure}[h!]
\begin{center}
\includegraphics[width=8.5cm]{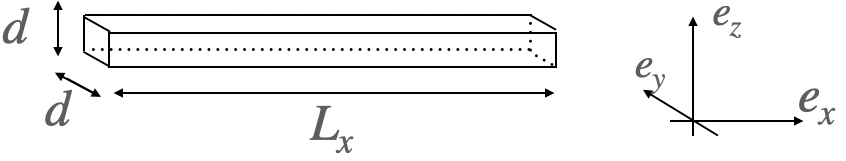}
\includegraphics[width=7cm]{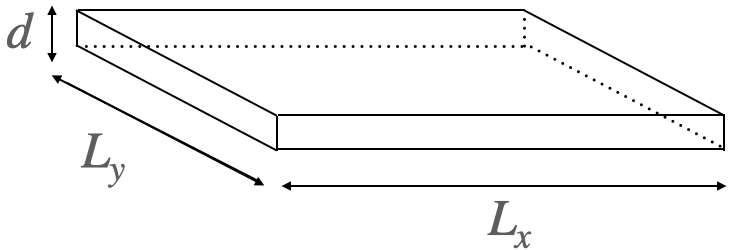}
\caption{Illustration of the two generic geometries corresponding to a 1D nanowire (left) and a 2D nanolayer (right)}
 \label{fig_geometry}
\end{center}
\end{figure}

The stochastic LLG equation \eqref{LL_stocha} is solved numerically using a \textmd{Python} code.\footnote{Available at: \url{https://gitlab.math.unistra.fr/llg3d/llg3d}} 
The simulations are performed on the time interval $[0, t_f]$, with $t_f$ the final time, and on a 3D finite domain $[0,L_x]\times[0,L_y]\times[0,L_z]$. To mimic a 1D nanowire, this numerical domain is taken with a small square cross-section in the $(\boldsymbol{e}_y, \boldsymbol{e}_z)$ plane: $[0,L_x]\times[0,d]\times[0,d]$, with $d\ll L_x$. To mimic a 2D nanolayer, the 3D domain is taken with a small thickness in the $\boldsymbol{e}_z$ direction: $[0,L_x]\times[0,L_y]\times[0,d]$, with $d\ll L_x, L_y$. Figure~\ref{fig_geometry} illustrates those two geometries.
All numerical parameters are listed in Table~\ref{table_numerical_parameters}. 

\begin{table}
\begin{center}
\begin{tabular}{p{3cm}p{5cm}p{7.5cm}}
\toprule
\multicolumn{3}{c}{\textbf{Discretization time and space parameters}}\\
\hline
$\Delta t$& time step& $10 $ fs \\
$t_f$& final time of the simulation & $50$ ps\\
$N$ & number of time iterations& 5000 \\
$n$ & time index & $0\leq n\leq N$\\ 
&&\\
$\Delta x=\Delta y=\Delta z$ & space size of the meshes & $1.0 $ nm \\
$d$ & small size parameter & $11.0 $ nm $\leq d\leq 41.0$ nm\\
$L_x, L_y, L_z$ &  size of the ferromagnetic ob- & nanowire ($L_x=6000$ nm, $L_y=L_z=d$ nm)\\
&ject& nanolayer ($L_x=L_y=600$ nm, $L_z=d$ nm) \vspace*{0.1cm}\\
$J_x, J_y, J_z$ & number of mesh points & nanowire ($J_x=6000$, $J_y=J_z=\lfloor d\rfloor$)\\
&& nanolayer ($J_x=J_y=600$, $J_z=\lfloor d\rfloor$) \vspace*{0.1cm}\\
$i, j, k$ & space  indices & $0\leq i\leq J_x, 0\leq j\leq J_y, 0\leq k\leq J_z$\\ 
&&\\
\bottomrule
\end{tabular}
\begin{tabular}{p{1.5cm}p{6cm}p{8cm}}
\multicolumn{3}{c}{\textbf{Numerical variables}}\\
\hline
$\boldsymbol{m}^0$ & initialization of the magnetic moments & $\begin{pmatrix}1\\0\\0\end{pmatrix}$ for all points $(x_i, y_j, z_k)$\vspace*{0.1cm}\\
$\overline{m}_1(t)$ & spatial average of one realization & $\displaystyle \frac{1}{L_xL_yL_z}\int_{[0,L_x]\times[0,L_y]\times[0,L_z]} m_1(t,\boldsymbol{x})d\boldsymbol{x}$\vspace*{0.2cm}\\
$\tau$ & transient time behavior of $\overline{m}_1(t)$&40 ps \vspace*{0.1cm}\\
$M_{\text{tot}}$ &  \begin{tabular}[t]{@{}l@{}} total magnetization \\ \begin{small}(spatial and stochastic averages at $t_f$) \end{small} \end{tabular} & $ \displaystyle \frac{1}{t_f-\tau}\int_{\tau}^{t_f} \frac{1}{L_xL_yL_z}\int_{[0,L_x]\times[0,L_y]\times[0,L_z]} m_1(t,\boldsymbol{x})d\boldsymbol{x}dt$\\
%& \begin{small}(spatial and stochastic averages at $t_f$)\end{small} & %\vspace*{0.2cm}\\
$t_{\text{conv}}$ & convergence time for $\overline{m}_1(t)$ to reach its plateau state $M_{\text{tot}}$ &$\underset{t\in[0, t_f]}{\text{inf}}  |\overline{m}_1(t)-M_{\text{tot}}|<0.1$\\
\bottomrule
\end{tabular}
\caption{Numerical parameters used in the \textmd{Python} code}
\label{table_numerical_parameters}
\end{center}
\end{table}

\paragraph{Discretization.}
Numerically, let $\Delta t>0$ and $\Delta x=\Delta y=\Delta z>0$ be the time step and the space steps in each space direction, respectively. We define time-discrete and space-discrete points with $N=\lfloor \frac{t_f}{\Delta t}\rfloor$, $J_x=\lfloor \frac{L_x}{\Delta x}\rfloor$, $J_y=\lfloor \frac{L_y}{\Delta x}\rfloor$, $J_z=\lfloor \frac{L_z}{\Delta x}\rfloor$

$$t^n=n\Delta t,\ \ \ \ \ 0\leq n\leq N$$ and $$(x_i, y_j, z_k)=(i\Delta x, j\Delta x, k\Delta x), \ \ \ \ 0\leq i\leq J_x, \ \ 0\leq j\leq J_y, \ \ 0\leq k\leq J_z.$$
The numerical solution $\boldsymbol{m}_{i,j,k}^n$ approximates the exact one $\boldsymbol{m}(t^n, x_i, y_j, z_k)$ on each discrete point. 

As the LLG equation \eqref{LLGeq} is valid at mesoscopic -- and not atomistic -- length scales,
the spatial steps $\Delta x$, $\Delta y$, $\Delta z$ are each much larger than the lattice constant $a_{\text{eff}}$, and the continuous magnetic moment vector filed $\boldsymbol{m}(t,\boldsymbol{x})$ actually represents an average of the atomic spins over a volume $\Delta x \Delta y \Delta z$.

\paragraph{Time evolution: Heun method.} For consistency with the continuous problem, the stochastic LLG equation \eqref{LL_stocha} must be discretized using a numerical method whose solution converges to the Stratonovich continuous solution. For this purpose, the modified Heun method \cite{Ruemelin1982} is chosen for the time integration and a second-order finite difference method is used for the discretization of the Laplacian operator.
 
Following \cite{Klughertz_these}, for further simplicity, we rewrite Eq.~\eqref{LL_stocha}  as 
\begin{equation}\label{LL_stocha_reecrite}
d\boldsymbol{m}= \boldsymbol{F}(\boldsymbol{m}, t)dt + \sum_{j\in\{1, 2, 3\}}\boldsymbol{G}_{j}(\boldsymbol{m})\,\nu \, dW_j(t) ,
\end{equation}
 with \begin{equation*}
 \boldsymbol{F}(\boldsymbol{m}, t) = -\gamma_0\boldsymbol{m}\times\boldsymbol{H}_{\text{eff}}-\gamma_0\alpha\,\boldsymbol{m}\times\left(\boldsymbol{m}\times\boldsymbol{H}_{\text{eff}}\right)
 \end{equation*}
 being the deterministic part and $\boldsymbol{G}_{j} = \begin{pmatrix}G_{1,j}\\G_{2,j}\\G_{3,j}\end{pmatrix}$ the factor term of the stochastic process, with 
 \begin{equation*}
 G_{i,j}=\gamma_0 m_k\epsilon_{i,j,k} - \gamma_0 \alpha (m_im_j-\delta_{ij}),
 \end{equation*} where $\epsilon_{i,j,k}$ is the Levi-Civita symbol.

After initializing the magnetization at the initial time $t=0$: $\boldsymbol{m}^0_{i,j,k}=\begin{pmatrix}m_1\\m_2\\m_3\end{pmatrix}^0_{i,j,k}=\boldsymbol{m}(0, x_i, y_j, z_k)$, the (stochastic) Heun method consists in the following steps to go from the time step $n$ to the time step $n+1$:
\begin{itemize}
\item Generate a random vector $\boldsymbol{R}^n\in\mathbb{R}^3$ according to a reduced centered normal distribution, using a pseudo-random number generator. Define $\Delta \boldsymbol{W}^n = \sqrt{\Delta t} \boldsymbol{R}^n$;
\item Compute $\nu_{\text{num}}$, the numerical version of the standard deviation $\nu$ (see Eq.~\eqref{eq_variance_num});
\item Define $ \boldsymbol{\kappa}_1 = \boldsymbol{F}(\boldsymbol{m}^n, t^n)$ and $ \boldsymbol{s}_{1, j}=\boldsymbol{G}_{j}(\boldsymbol{m}^n)$;
\item Define $ \boldsymbol{\kappa}_2 = \boldsymbol{F}(\boldsymbol{m}^n+\Delta t\boldsymbol{\kappa}_1 + \sum_{j\in\{1, 2, 3\}} \boldsymbol{s}_{1, j}\nu_{\text{num}}\Delta W_j^n, t^{n}+\Delta t)$ and $ \boldsymbol{s}_{2, j}=\boldsymbol{G}_{j}(\boldsymbol{m}^n+\Delta t\boldsymbol{\kappa}_1+ \sum_{j\in\{1, 2, 3\}} \boldsymbol{s}_{1, j}\nu_{\text{num}}\Delta W_j^n)$;
\item Update $\boldsymbol{m}^{n+1}=\boldsymbol{m}^n+\left(\frac{1}{2} \boldsymbol{\kappa}_1 + \frac{1}{2} \boldsymbol{\kappa}_2 \right)\Delta t + \sum_{j\in\{1, 2, 3\}}\left(\frac{1}{2} \boldsymbol{s}_{1, j} + \frac{1}{2} \boldsymbol{s}_{2, j} \right)\nu_{\text{num}}\Delta W_j^n$;
\item Renormalize the magnetic moment: $\boldsymbol{m}^{n+1}=\frac{\boldsymbol{m}^{n+1}}{\|\boldsymbol{m}^{n+1}\|}$, so that it remains on the unit sphere $\mathbb{S}^2$.
\end{itemize}

A mid-point numerical technique would be also possible alternative to the Heun method, see \cite{dAquino_et_al, Ragusa_et_al}.

The choice of a time-explicit discretization of the Laplacian operator induces a restrictive condition on the time step $\Delta t$ and the space step $\Delta x$ to ensure the stability of the scheme: $\Delta t \lesssim \Delta x^2/2$ (Courant-Friedrichs-Lewy condition).

At the domain boundaries, we take the usual Neumann condition:
$\partial \boldsymbol{m}/\partial \boldsymbol{n}=0$,
where $\boldsymbol{n}$ is the outgoing normal vector.

\section{Numerical code validation}\label{sec:validation_code}

In the forthcoming simulations, three types of ferromagnetic materials will be considered: (i) nickel with a FCC lattice and cubic anisotropy, (ii) iron with a BCC lattice and a cubic anisotropy, and (iii) cobalt with a HCP lattice and uniaxial anisotropy. For each case, we shall study both 1D nanowires and 2D nanolayers.
All physical parameters are listed in Table~\ref{table_constantes_physiques}.

Here, we will perform several tests to validate our numerical code. In the following sections, we will focus on the scaling of the Curie temperature with the size of the system, for each type of nano-object. 
We recall that our code relies on the time-dependent LLG equation. We solve numerically this equation with a fully magnetized initial condition, where $\boldsymbol{m}(t=0)$ is uniform and directed along the $\boldsymbol{e}_x$ axis, and a given temperature. Then, we wait that magnetic moment relaxes to a lower value under the effect of the temperature and determine its value by averaging over a sufficiently long period of time.

\GMcorr{
The forthcoming test cases show that the present method is capable of reproducing the correct Curie temperature with a relative accuracy of about $5\%$, despite the simplicity of the underlying model and, most importantly, the fact that it does not depend on any free adjusting parameter. This is close to the accuracy obtained with more elaborate and computationally demanding atomistic models \cite{Evans2015,Evans2014,Hovorka2012}. 
In addition, the latter can only deal with small nanostructures, with typical size $\lesssim 10^3\, \rm nm^3$ (see, e.g., \cite{Hovorka2012}), or  $20^3\, \rm nm^3$ but using periodic boundary  conditions, see \cite{Evans2015}. In contrast, we could push our calculations up to $50^3\, \rm nm^3$ for cubic structures or $600^2 \times 40 \, \rm nm^3$ for 2D films.
We also note that, without the temperature scaling adopted here (see Sec.~\ref{sec:Tscaling}), the error on $T_{\rm C}$ would be much higher, possibly one order of magnitude or so \cite{Grinstein_Koch_2003}.
}

\subsection{Test-case details} \label{sec:testdetails}
Except for the following Sec.~\ref{sec:independence} -- where the independence of the results on the space and time discretizations are tested on a 3D cube --  all  numerical simulations are preformed on three ferromagnetic materials (cobalt, iron and nickel) and two geometries (see Figure~\ref{fig_geometry}):
\begin{itemize}
\item 1D nanowire with small square cross-section in the $(\boldsymbol{e}_y, \boldsymbol{e}_z)$ plane: $[0,L_x]\times[0,d]\times[0,d]$ with $L_x=6000$ nm and $11$ nm $\leq d\leq 41$ nm;
\item 2D nanolayer with small thickness in the $\boldsymbol{e}_z$ axis: $[0,L_x]\times[0,L_y]\times[0,d]$ with $L_x=L_y=600$ nm and $11$ nm $\leq d\leq 41$ nm.
\end{itemize}
The numerical parameters are always fixed to (see Table~\ref{table_numerical_parameters}):
\begin{equation*}
\Delta t =  10 \; \text{fs}, \quad t_f =50 \; \text{ps}, \quad N = 5000, \quad
\Delta x=\Delta y=\Delta z = 1 \; \text{nm}.
\end{equation*}
The initialization of the magnetic moments $\boldsymbol{m}^0$ is chosen uniform and directed along the $\boldsymbol{e}_x$ axis in all test cases: $\boldsymbol{m}^0_{i,j,k}=\begin{pmatrix}1\\0\\0\end{pmatrix}$ for all $i,j,k$. 

With the choice of initialization, the magnetization is initially equal to 1 (when the magnetic moments are all aligned and have a norm equal to 1), then falls to zero at the Curie temperature (when the magnetic moments are randomly aligned  due to the thermal noise). Since Eq.~\eqref{LL_stocha} is stochastic, the average of the magnetic moments $\boldsymbol{m}$ over the entire ferromagnetic domain may differ from one realization to another, so this average should be calculated over several realizations. Since the initialization is oriented along the $\boldsymbol{e}_x$ axis and in the absence of any external magnetic field, the average of $\boldsymbol{m}$ along this direction, i.e. $m_1$, is enough to characterize the magnetic state of the system. Hence, we define the total (in space) magnetization
\begin{equation}\label{magnet_totale_1}
M_{\text{tot}} = \left\langle \frac{1}{L_xL_yL_z}\int_{[0,L_x]\times[0,L_y]\times[0,L_z]} m_1(t_f,\boldsymbol{x})d\boldsymbol{x} \right\rangle,
\end{equation}
with $\langle\cdot\rangle$ denoting the statistical average over many realizations. 
In order to simplify this calculation, we assume that the stochastic process of Eq.~\eqref{LL_stocha} is ergodic, so that the statistical average may be replaced by the time average for a single, sufficiently long realization. In practice, we plot several realizations, look at the time $\tau$ after which the transient regime gives way to the stationary regime, and finally take the time average from this {\em transient time} $\tau$ up to the final time $t_f$. Thus, the total magnetization is now defined as 
\begin{equation}\label{magnet_totale_2}
M_{\text{tot}} =\frac{1}{t_f-\tau}\int_{\tau}^{t_f} \frac{1}{L_xL_yL_z}\int_{[0,L_x]\times[0,L_y]\times[0,L_z]} m_1(t,\boldsymbol{x})d\boldsymbol{x}dt = \frac{1}{t_f-\tau}\int_{\tau}^{t_f} \overline{m}_1(t) dt.
\end{equation}

\begin{figure}[h!]
\centering
\begin{tabular}{cc}
\includegraphics[width=8cm]{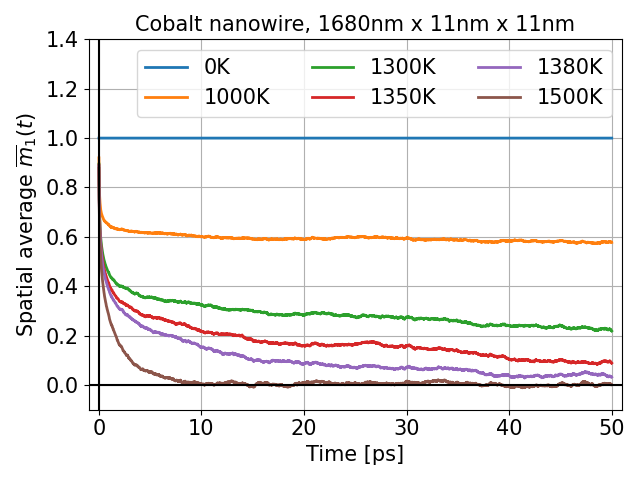}
&
\includegraphics[width=8cm]{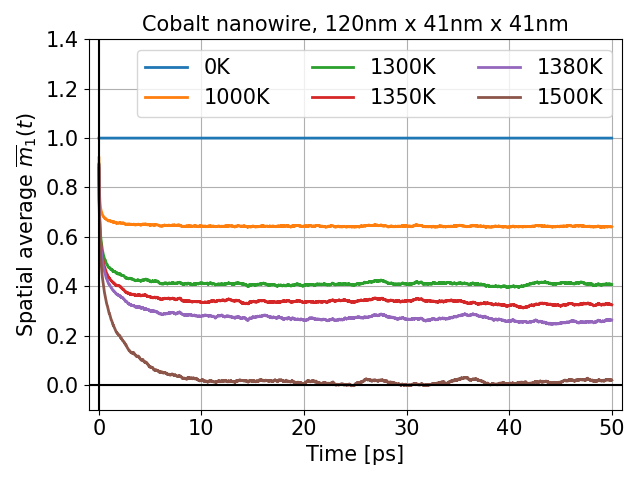} \\
\includegraphics[width=8cm]{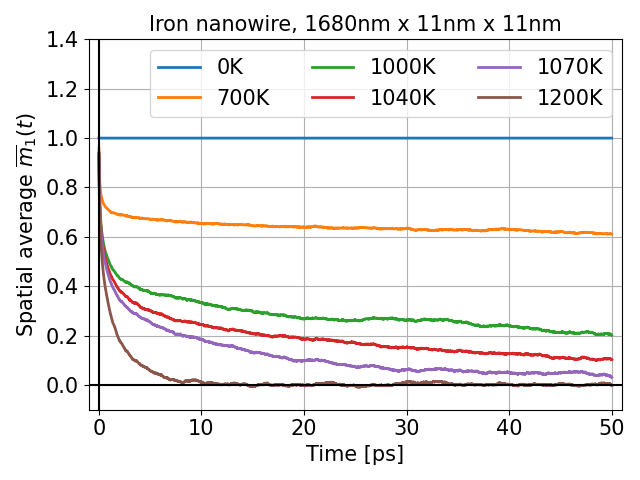}
&
\includegraphics[width=8cm,height=6cm]{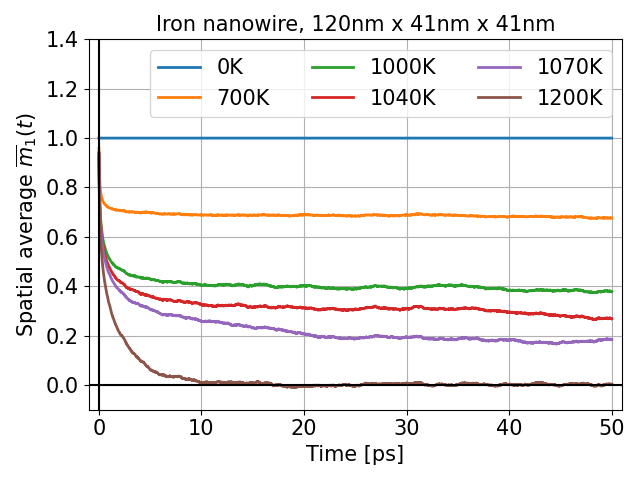}\\
\includegraphics[width=8cm]{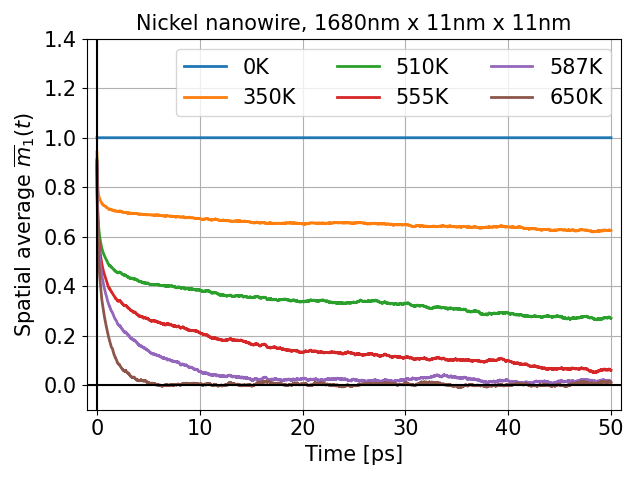}
&
\includegraphics[width=8cm,height=6cm]{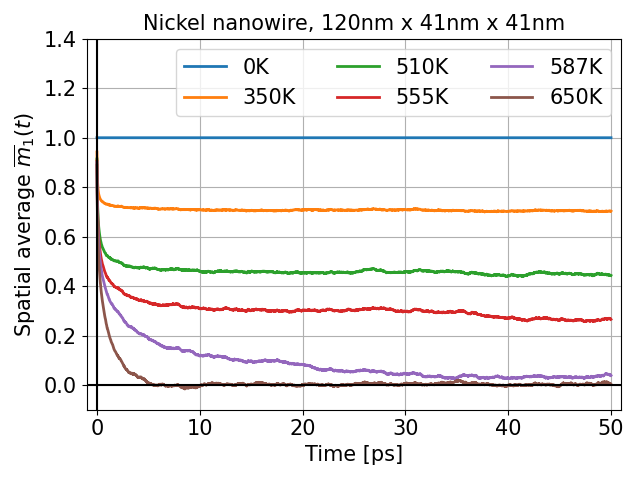}
\end{tabular}
\caption{Spatially averaged $x$ component of the magnetic moment $\overline{m}_1(t)$ for one statistical realization, as a function of time $t$ for a ferromagnetic nanowire. Colors correspond to different temperatures, as indicated on the figure. Top panels: cobalt; middle panels: iron; bottom panels: nickel. The left column corresponds to nanowires with dimensions (in nm): $1680\times 11\times 11$, the right column to nanowires with dimensions (in nm): $120 \times 41\times 41$. 
\label{fig_magnetization_en_fct_temps}}
\end{figure}

Figure~\ref{fig_magnetization_en_fct_temps} illustrates the spatial average of the $x$ component of the magnetic moment  $\overline{m}_1(t)=\frac{1}{L_xL_yL_z}\int_{[0,L_x]\times[0,L_y]\times[0,L_z]} m_1(t,\boldsymbol{x})d\boldsymbol{x}$ for one statistical realization, for cobalt (top), iron (middle) and nickel (bottom). Two geometries of nanowire are tested: $[0,1680]\times[0,11]\times[0,11]$ with $\Delta x=1$ nm (left column) and $[0,120]\times[0,41]\times[0,41]$ with $\Delta x=1$ nm (right column). Other numerical parameters $\Delta t$, $t_f$ and $N$ are defined in Table~\ref{table_numerical_parameters}. Different temperatures are used (represented with different colors) and lead to the same conclusion for all test cases: $\tau=0.8\,t_f = 40$ ps  is a correct choice both for reaching the stationary state and for having enough time left for a representative average (this choice represents a time average over the last 1000 time steps, and $t_f-\tau=10$ ps in Eq.~\eqref{magnet_totale_2}).

In practice, the Curie temperature is determined numerically by plotting $M_{\text{tot}}$ as a function of the temperature $T$, and defining $T_{\text{C}}$ as the temperature for which $M_{\text{tot}}$ is smaller than a certain threshold, fixed to 0.1, i.e., $T_{\text{C}} := \underset{T}{\text{argmin}\ } M_{\text{tot}}(T) < 0.1$.

\subsection{Dependence on the numerical parameters}\label{sec:independence}
Here, we show that our results do not depend on either the time step $\Delta t$ or the computational cell size $\Delta x$.

\begin{figure}[h!]
\begin{center}
\includegraphics[width=8cm]{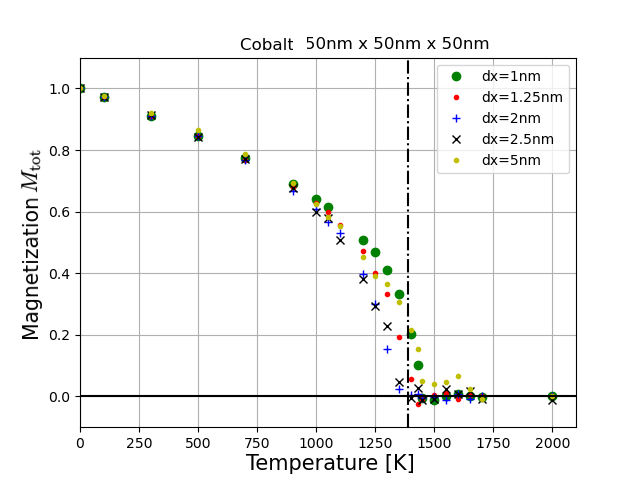}
\includegraphics[width=8cm]{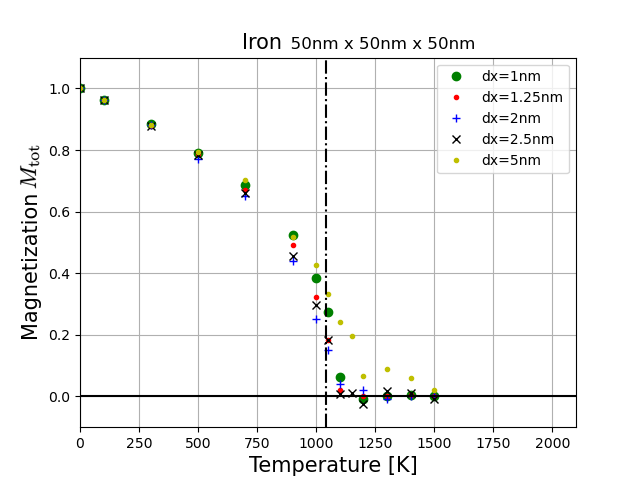}
\includegraphics[width=8cm]{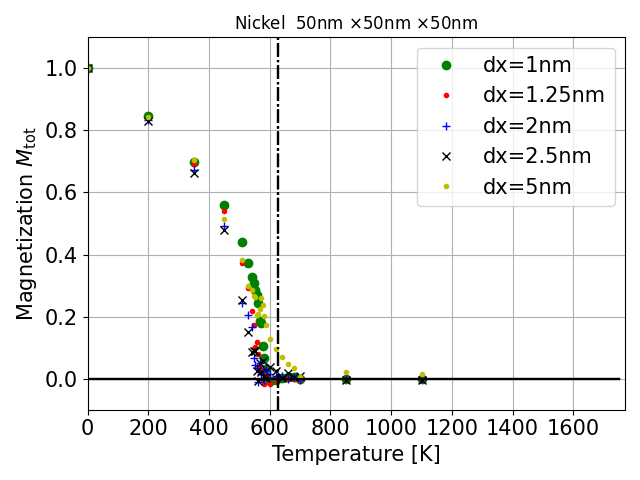}
\caption{Total magnetization $M_{\text{tot}}$, from Eq. \eqref{magnet_totale_2}, as a function of the temperature, for cubic cobalt (left panel), iron (right panel) and nickel (middle panel) nano-objects with dimensions $\rm 50~nm\times 50~nm\times 50~nm$. The different symbols and colors stand for different computational cell sizes $\Delta x$, going from 1~nm to 5~nm. The black vertical dash-dotted lines represent the bulk Curie temperatures as given in Table~\ref{table_constantes_physiques}. 
\label{fig_independant_dx}}
\end{center}
\end{figure}

\begin{figure}[h!]
\begin{center}
\includegraphics[width=8cm]{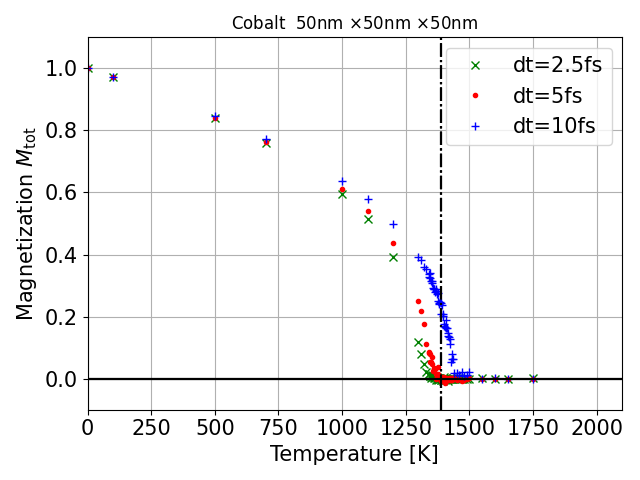}
\includegraphics[width=8cm]{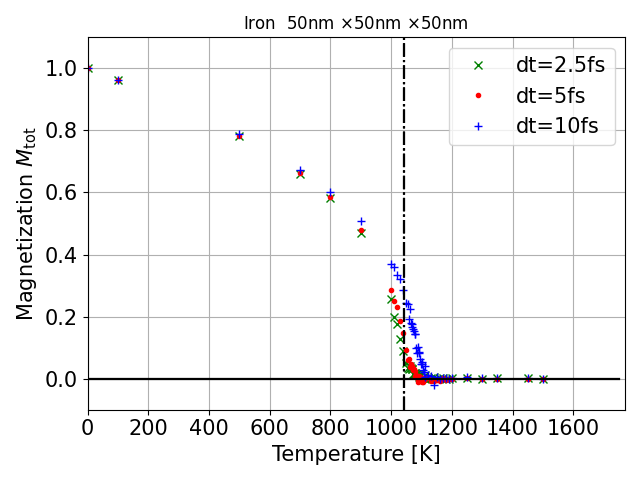}
\includegraphics[width=8cm]{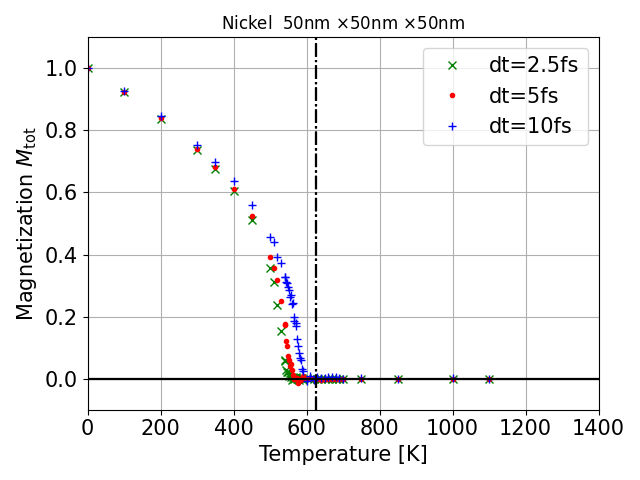}
\caption{Total magnetization $M_{\text{tot}}$, from Eq. \eqref{magnet_totale_2}, as a function of temperature, for a cobalt (left), iron (right) or nickel (middle) nanowire with dimensions 50~nm$\times$50~nm$\times$50~nm. The different symbols stand for different time steps $\Delta t =2.5~\rm fs$ (green crosses), 5~fs (red circles), and 10~fs (blue crosses). The computational grid size is $\Delta x=1$~nm. The black vertical dash-dotted lines represent the bulk Curie temperatures as given in Table~\ref{table_constantes_physiques}. \label{fig_independant_dt}}
\end{center}
\end{figure}

Figure~\ref{fig_independant_dx} shows the total magnetization $M_{\text{tot}}$ as a function of the temperature for different cell sizes, going from 1~nm to 5~nm, for a cubic object of dimensions $\rm 50~nm\times 50~nm \times 50~nm$, for cobalt, iron and nickel. All other parameters are identical ($\Delta t=10$ fs, $t_f=50$ ps, $N=5000$ and $\tau=40.0$ ps). The results are indeed independent on $\Delta x$, and the computed Curie temperatures are very close to the experimental values for the bulk materials (see Table~\ref{table_constantes_physiques}). A slight discrepancy starts occurring for iron at $\Delta x= 5~\rm nm$.

Figure~\ref{fig_independant_dt} shows the dependence of the numerical results with respect to the time step $\Delta t$, again for cubic nano-objects of side 50 nm, with computational cell size $\Delta x=1$ nm. The final time is always $t_f=50$ ps, so that the number of time steps is $N=5\times 10^{3}$ for $\Delta t=10$ fs (blue curve), $N=10^{4}$ for $\Delta t =5$ fs (red curve), and $N=2\times 10^{4}$ for $\Delta t=2.5$ fs (green curve). According to the value of $\Delta t$, the time-averaged $M_{\text{tot}}$ in Eq.~\eqref{magnet_totale_2} includes the last 1000, 2000 or 4000 time iterations. The computed Curie temperature varies only slightly with the time step, and appears to have converged for $\Delta t= 5~\rm fs$.

\GMcorr{However, the results are already quite satisfying, albeit not perfect, for $\Delta t= 10~\rm fs$. For this reason, and since we need to perform a large number of runs in order to deduce the scaling with size (see Sec.~\ref{sec:size_effects}), all forthcoming simulations will be performed with $\Delta t= 10~\rm fs$. This value is in line with the earlier simulations of Hahn \cite{Hahn_2019}.
One should further keep in mind that, without the temperature scaling adopted here and detailed in Sec.~\ref{sec:Tscaling}, the error on $T_{\rm C}$ would be much higher, possibly one order of magnitude \cite{Grinstein_Koch_2003}. The values we obtain are thus a good trade-off between accuracy and computational efficiency. 
We also note that the experimental bulk values of $T_{\rm C}$ displayed on Fig. ~\ref{fig_independant_dt} as vertical lines are given for reference, but never used in practice to determine the size effects. Instead, the effective "bulk" values of $T_{\rm C}$ that we use are those obtained computationally with the largest structure that we consider (see Sec.~\ref{sec:size_effects} for details).  
}

\subsection{Magnetization curve: Bloch's law and Curie's law} \label{sec:bloch}

In this section, we check that the numerically calculated magnetization $M_{\text{tot}}(T)$ satisfies the standard Bloch's and Curie's laws, respectively at low temperatures $T \ll T_{\text{C}}$ and near the Curie temperature, $T \lesssim T_{\text{C}}$.
As we have already verified that the spatial and temporal steps do not influence the result, $\Delta t$ and $\Delta x$ will be fixed as specified in Table ~\ref{table_numerical_parameters}, i.e. $\Delta t=10$ fs and $\Delta x=1$ nm.

\begin{figure}[h!]
\begin{center}
\includegraphics[width=8cm]{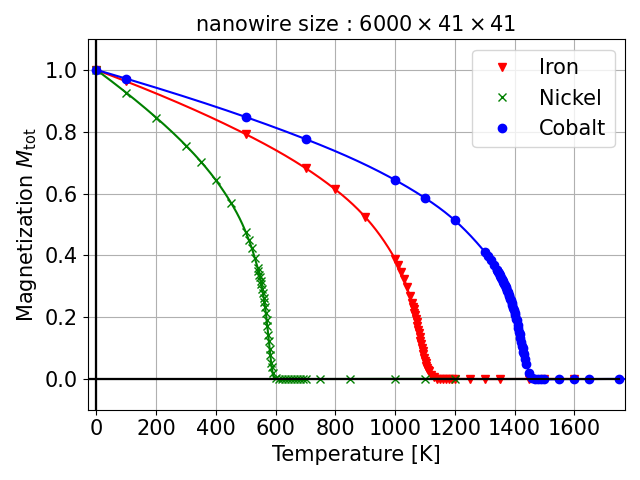}
\includegraphics[width=8cm]{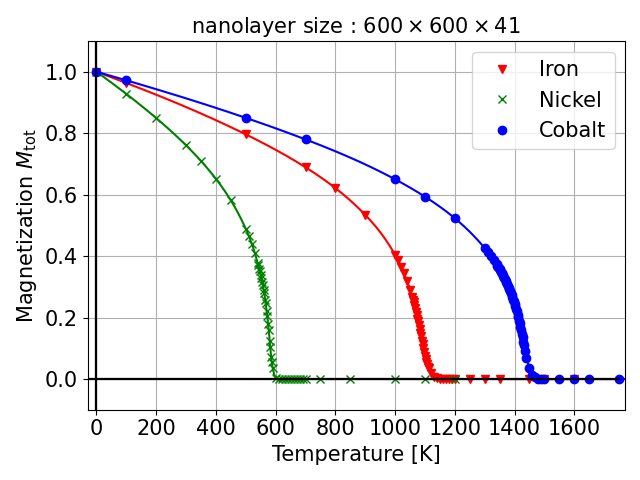}
\caption{Total magnetization $M_{\text{tot}}$ \eqref{magnet_totale_2} with respect to temperature with numerical parameters detailed in Table~\ref{table_numerical_parameters}. The Curie temperature corresponds to the first temperature at which magnetization falls to zero. Simulation results are represented by dots, the solid curves are an interpolation based on cubic splines. \label{fig_curie_temp}}
\end{center}
\end{figure}

Figure~\ref{fig_curie_temp} illustrates the behavior of the magnetization curve  $M_{\text{tot}}(T)$ for the two geometric configurations considered here (nanowires and nanolayers), for iron (red triangles), cobalt (blue circles) and nickel (green crosses). Results are in broad agreement with the expected magnetization curves, and the computed Curie temperatures are close to the experimental values found in the literature for bulk materials \cite{Kittelbook2018}, see  Table~\ref{table_constantes_physiques}.

\begin{figure}[h!]
\begin{center}
\includegraphics[width=8cm]{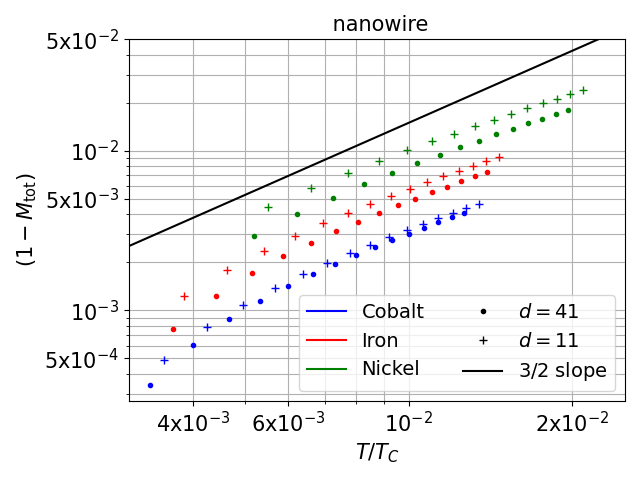}
\includegraphics[width=8cm]{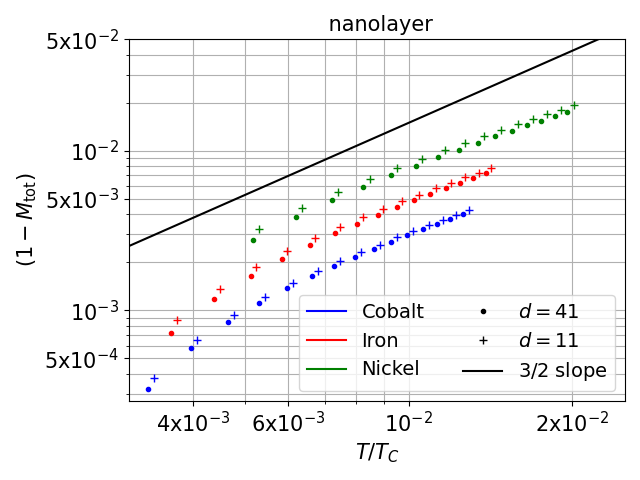}
\caption{Bloch's law. Behavior of $1-M_{\text{tot}}(T)$ as a function of $T/T_{\text{C}}$ in logarithmic scale.  The materials are represented by different colors: iron (red), cobalt (blue) and nickel (green). The left panel corresponds to nanowires and the right panel to nanolayers, with sizes $d=11 \, \rm nm$ (crosses) and $d=41 \, \rm nm$ (dots). Each curve was multiplied by a multiplicative factor in the $y$-axis for easier reading (1 for cobalt, 1.6 for iron and 2.5 for nickel).
The black solid lines represent the theoretical $3/2$ slope.}
\label{fig_loi_Bloch}
\end{center}
\end{figure}

Bloch's law states that total magnetization $M_{\text{tot}}(T)$, for low temperatures, behaves as follows: 
\begin{equation*}
M_{\text{tot}}(T) \underset {T\rightarrow 0}{\sim} 1-\left(\frac{T}{T_{\text{C}}}\right)^{3/2},
\end{equation*}
which can be rewritten as: $\log(1-M_{\text{tot}}) \sim  \frac{3}{2}\log\left(T/T_{\text{C}} \right)$.
Figure~\ref{fig_loi_Bloch} checks this behaviour on a log-log scale, for a nanowire (left) and a nanolayer (right), with sizes corresponding to the two extreme cases in Table \ref{table_temperature_curie}, i.e. $d=11 \, \rm nm$ and 41~nm. 
The theoretical $3/2$ slope is represented as a solid black line and matches the numerical results  quite well.

\begin{figure}[h!]
\begin{center}
\includegraphics[width=8cm]{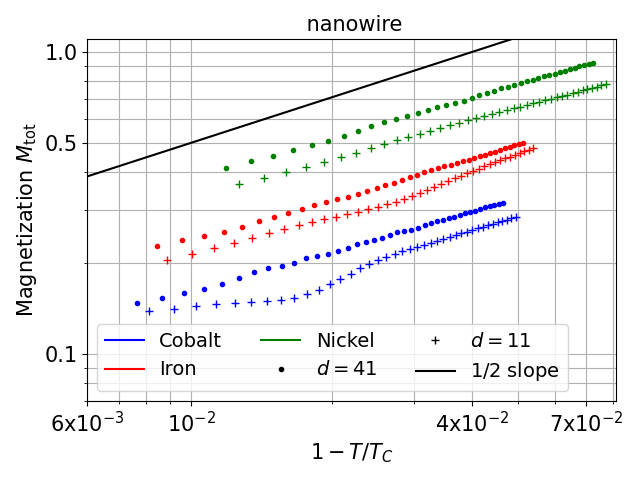}
\includegraphics[width=8cm]{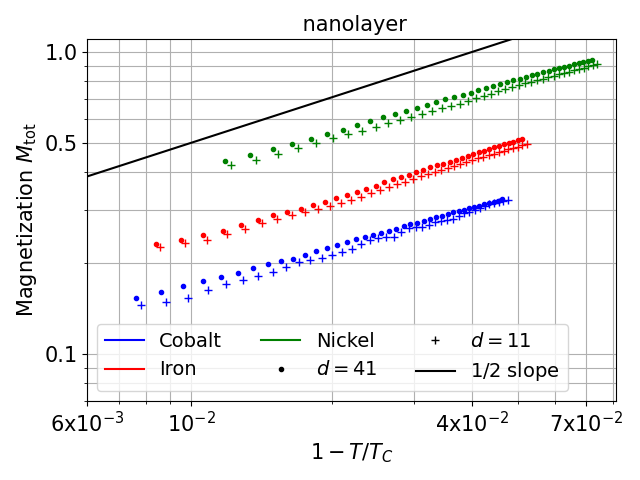}
\caption{Curie's law. Behavior of $M_{\text{tot}}(T)$ as a function of $1-T/T_{\text{C}}$ in logarithmic scale.  The materials are represented by different colors: iron (red), cobalt (blue) and nickel (green). The left panel corresponds to nanowires and the right panel to nanolayers, with sizes $d=11 \; \rm nm$ (crosses) and $d=41 \; \rm nm$ (dots). Each curve was multiplied by a multiplicative factor in the $y$-axis for easier reading (1 for cobalt, 1.6 for iron and 2.5 for nickel). 
The black solid lines represent the theoretical $1/2$ slope.} 
\label{fig_magnetization_exponent}
\end{center}
\end{figure}

Next, we check the Curie law, valid near $T_{\text{C}}$:
\begin{equation*}
M_{\text{tot}}(T) \underset{T\underset{<}{\rightarrow}T_{\text{C}}}{\sim} \left(1-\frac{T}{T_{\text{C}}}\right)^{1/2},
\end{equation*}
which may be rewritten as: $\log(M_{\text{tot}}) \sim \frac{1}{2}\log\left(1-\frac{T}{T_{\text{C}}}\right)$. Figure~\ref{fig_magnetization_exponent} shows the behavior of the magnetization $M_{\text{tot}}$ as a function of $1-T/T_{\text{C}}$ in logarithmic scale, for the same cases as those of Figure \ref{fig_loi_Bloch}. Again, the numerical results match rather well the theoretical $1/2$ slope.\footnote{\GMcorr{A visible deviation from linearity appears to occur for the Cobalt nanowire of size $d=11\, \rm nm$. We note that this is the smallest structure we considered, and thus the most subject to fluctuations. Indeed, closer inspection of Fig.~\ref{fig_magnetization_Tc_nanowire} (top right panel, blue curve ) reveals that the corresponding magnetization curve is noisier than the others. We have no detailed explanation of why this occurs for Cobalt and not for the other materials, but it may signal that we are reaching the size limit at which a micromagnetic description is appropriate. Also note that the correlation length for these structures is about 3~nm, hence not much smaller than their smallest size $d=11\, \rm nm$ (see Sec. \ref{sec:size_effects}).}
}

\section{Finite-size effects on the Curie temperature}\label{sec:size_effects}

This section is devoted to the study of the influence of size effect on the magnetization curve and Curie temperature for nanometric objects, both in 1D (nanowires) and 2D (nanolayers).
The objective is to vary the cross-section of the nanowire or the thickness of the nanolayer and study the variations induced in the Curie temperature. Throughout this section, numerical parameters are chosen as listed in Table~\ref{table_numerical_parameters}. 

\subsection{Size effects on the magnetization curve} 
\begin{figure}[h!]
\begin{center}
\includegraphics[width=8cm]{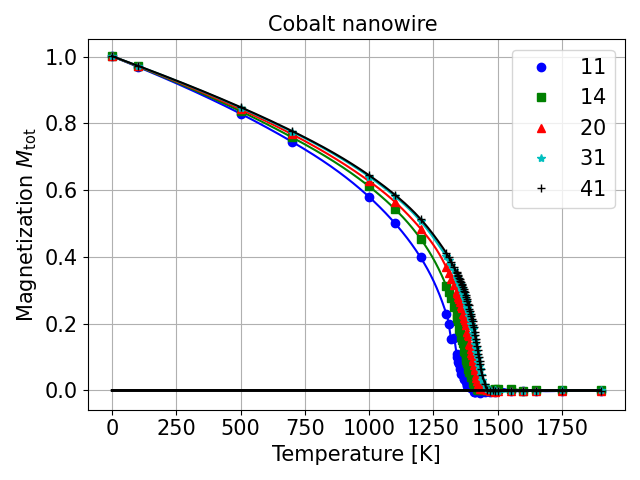}
\includegraphics[width=8cm]{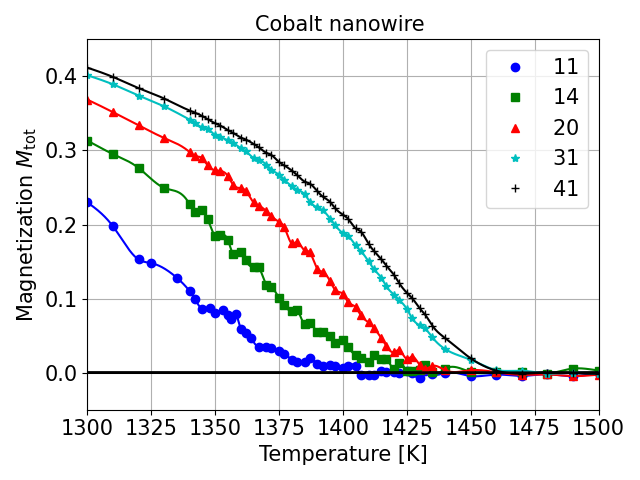}
\includegraphics[width=8cm]{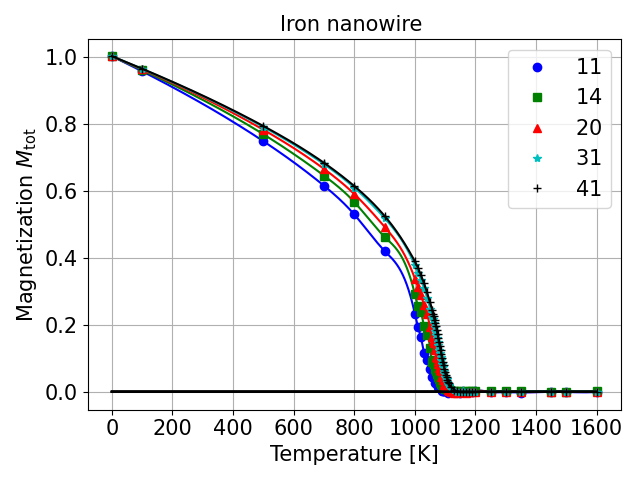}
\includegraphics[width=8cm]{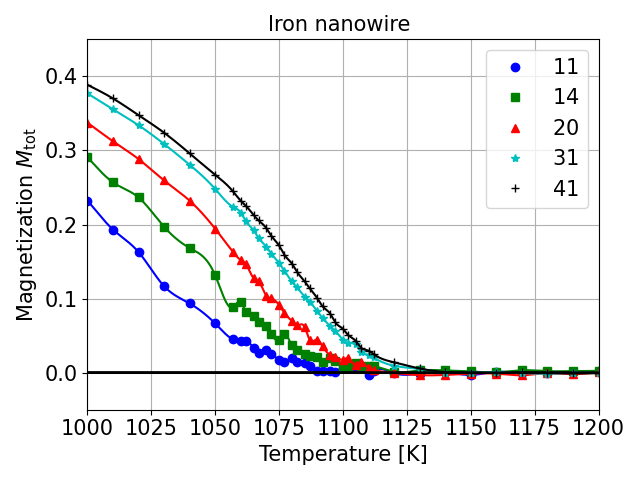}
\includegraphics[width=8cm]{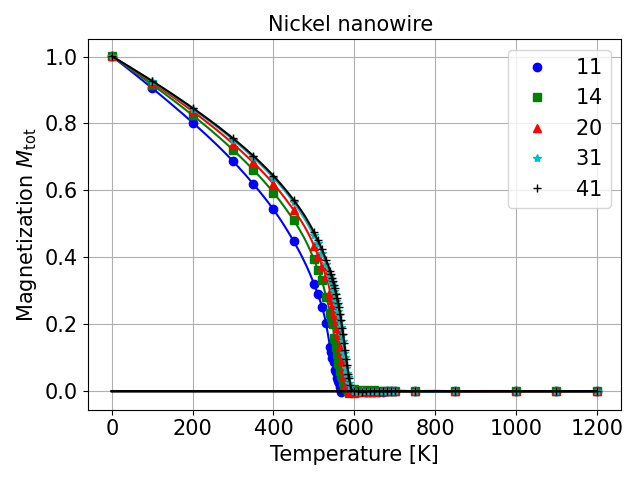}
\includegraphics[width=8cm]{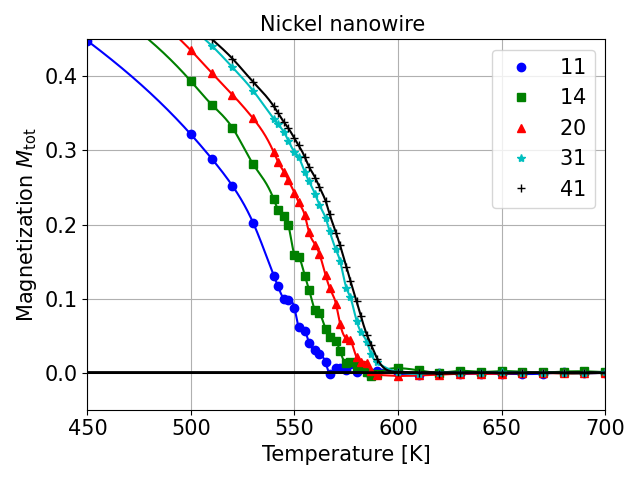}
\caption{Total magnetization $M_{\text{tot}}(T)$ as a function of the temperature for nanowires of different materials, with cross-section sizes going from $d=11\rm~ nm$ to $d=41\rm ~nm$. Top panels: cobalt; middle panels: iron; bottom panels: nickel. The right column is a zoom near the Curie temperature.  \label{fig_magnetization_Tc_nanowire}}
\end{center}
\end{figure}

\begin{figure}[h!]
\begin{center}
\includegraphics[width=8cm]{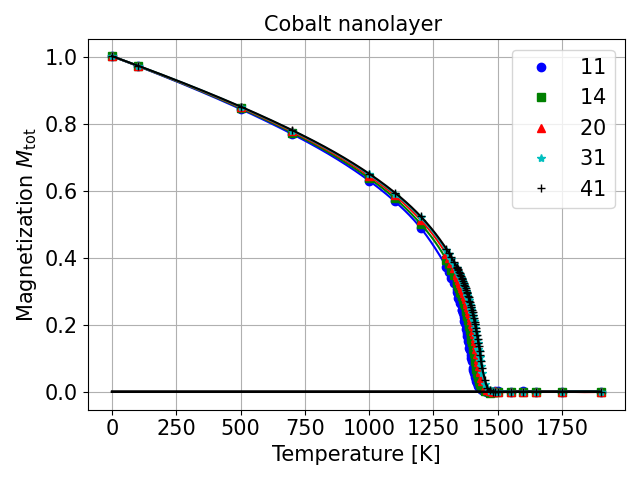}
\includegraphics[width=8cm]{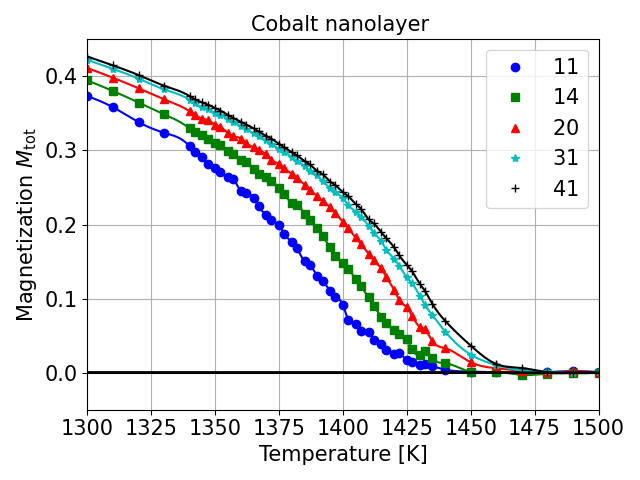}
\includegraphics[width=8cm]{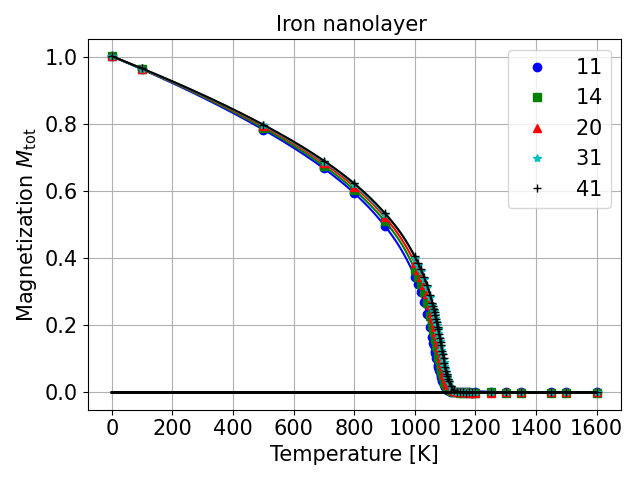}
\includegraphics[width=8cm]{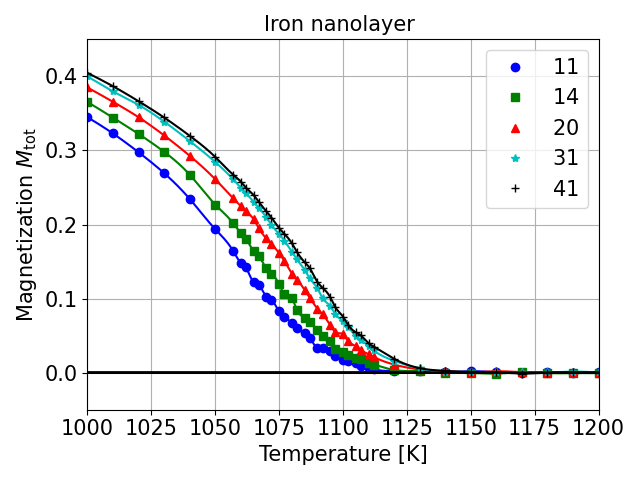}
\includegraphics[width=8cm]{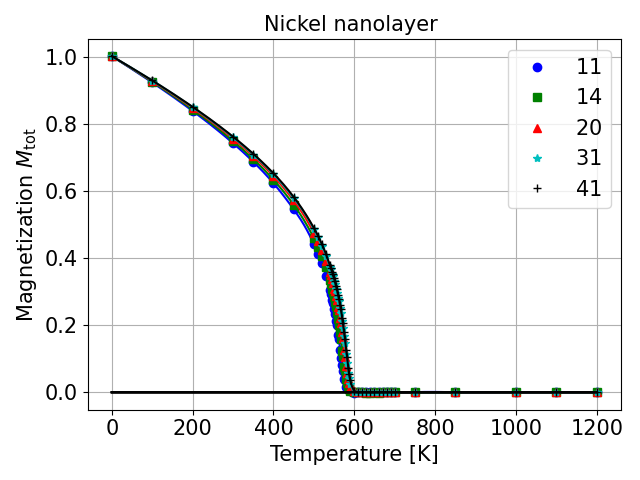}
\includegraphics[width=8cm]{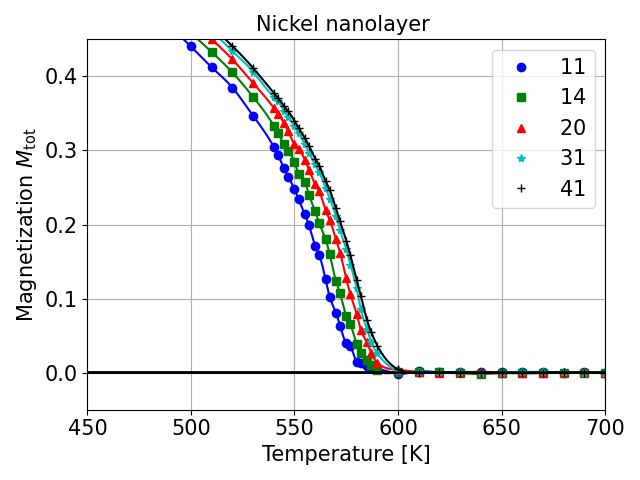}
\caption{Total magnetization $M_{\text{tot}}(T)$ as a function of the temperature for nanolayers of different materials, with thiknesses going from $d=11\rm ~nm$ to $d=41\rm ~nm$. Top panels: cobalt; middle panels: iron; bottom panels: nickel. The right column is a zoom near the Curie temperature. 
\label{fig_magnetization_Tc_nanolayer}}
\end{center}
\end{figure}

For all tested geometries, the computed Curie temperatures are close enough to the experimental values reported in the literature. This is achieved thanks to the scaling of the fluctuating thermal field as detailed in Sec.~\ref{sec:Tscaling}.
Nevertheless, we observe small variations with the size parameter $d$, which corresponds to the side of the square cross-section of a nanowire or the thickness of a nanolayer. Figure~\ref{fig_magnetization_Tc_nanowire} (for nanowires) and Figure~\ref{fig_magnetization_Tc_nanolayer} (for nanolayers) show the magnetization curves $M_{\text{tot}}(T)$, and illustrate how the Curie temperature increases with increasing size $d$. The computed values of $T_{\text{C}}$ are summarized in Table~\ref{table_temperature_curie}.
We also observe greater variability in $M_{\text{tot}}(T)$ for nanowires than for nanolayers. Size effects then appear to be stronger for lower-dimensional structures.

\begin{table}
\begin{center}
\begin{tabular}{p{1.2cm}p{1.8cm}p{1.8cm}p{1.8cm}||p{0.1cm}p{1.2cm}p{1.8cm}p{1.8cm}p{1.8cm}}
\toprule
\multicolumn{4}{c||}{\textbf{Nanowire $L_x\times d\times d$ with $L_x=6000$ nm}}&&\multicolumn{4}{c}{\textbf{Nanolayer $L_x\times L_y\times d$ with $L_x=L_y=600$ nm}}\\
$d$ [nm]&Cobalt&Iron&Nickel&&$d$ [nm]&Cobalt&Iron&Nickel\\
\hline
41&1427.26 &1090.05 & 579.62          && 41 &1433.91 &  1095.33 & 582.39\\
36&1423.74&1087.57 &  578.13          && 36 & 1432.20 & 1093.41 & 581.67\\
31& 1421.56& 1085.81 & 577.23         && 31 &1430.68 & 1092.13 &  581.01 \\
26 & 1414.62 & 1082.05 & 573.87        & & 26 & 1428.68 & 1091.49  & 579.81\\
21 & 1406.83 & 1076.85 & 571.05         && 21 &1422.79 &  1087.65 &   578.07\\
20 & 1401.23 & 1072.13 & 569.31        && 20 &  1421.75 &  1087.25 &  577.71\\
19 & 1397.04 & 1073.01 & 568.05        && 19 & 1421.27 & 1087.49 &  577.41 \\
18 & 1395.52 & 1069.65 & 566.91        && 18 &  1419.85& 1085.01 &  576.33 \\
17 & 1391.72 & 1069.17 & 564.45        && 17 & 1416.14 &  1083.41 & 575.61 \\
16 & 1385.55 & 1062.61 & 561.69       && 16 &  1415.10& 1082.05& 574.71 \\
15 & 1382.13 & 1062.45 & 560.67       && 15 & 1413.20 &  1080.53 &  573.63 \\
14 & 1375.19 & 1053.65 & 558.15      && 14 &  1410.35 & 1080.13 & 572.73  \\
13 &  1360.18 &  1045.89 &  552.69      && 13 & 1406.55 &  1076.37 &  570.87 \\
12 & 1352.11 & 1043.65 & 549.33         && 12 & 1400.37 &  1072.21 &  568.95\\
11 & 1342.04 & 1036.53 & 544.83         && 11 &  1398.09 &  1071.33 & 567.33 \\
\bottomrule
\end{tabular}
\caption{Curie temperatures (in Kelvin) obtained from the numerical simulations, for nanowires (left) and nanolayers (right) of different sizes and different materials.}
\label{table_temperature_curie}
\end{center}
\end{table}

\subsection{Power-law scaling of the Curie temperature}
Theoretical considerations \cite{Fisher1972} indicate that the Curie temperature should vary with the size $d$ following a power-law of the type:
\begin{equation}\label{loi_T_C_en_fct_d_calcul_lambda}
\frac{T_{\text{C}}(\infty)-T_{\text{C}}(d)}{T_{\text{C}}(\infty)}=\left(\frac{\xi_0}{d}\right)^{\lambda},
\end{equation}
where $\lambda$ is the critical exponent, $\xi_0$ is the correlation length at zero temperature, $T_{\text{C}}(\infty)$ is the Curie temperature of the bulk, and $T_{\text{C}}(d)$ the Curie temperature of a finite nano-object of size $d$. 
This type of power-law has been observed in many experiments \cite{Schneider1990,Huang1993,Li1992,Sun_2000,Zhang2001,Wang2011} and numerical simulations \cite{Almahmoud2010,Penny2019}.
Experimental works yielded  correlation length $\xi_0$ of the order of a few nanometers, with critical exponents in the range $\lambda \in [1-1.6]$. In contrast, an atomistic mean-field model \cite{Penny2019} yielded larger values, close to $\lambda = 2$ (range: $\lambda \in [1.82-2.17]$), for magnetite nanoparticles of different shapes and sizes. These values are also to be compared to those obtained from the 3D Heisenberg model ($\lambda=1.4$) and the 3D Ising model ($\lambda= 1.58$).

\begin{figure}[h!]
\begin{center}
\includegraphics[width=8cm]{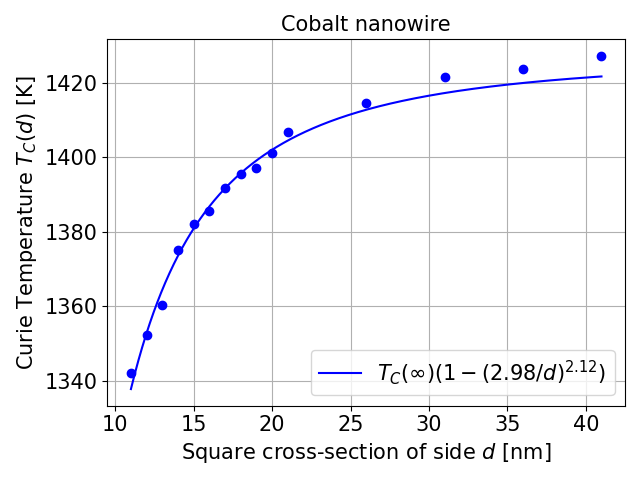}
\includegraphics[width=8cm]{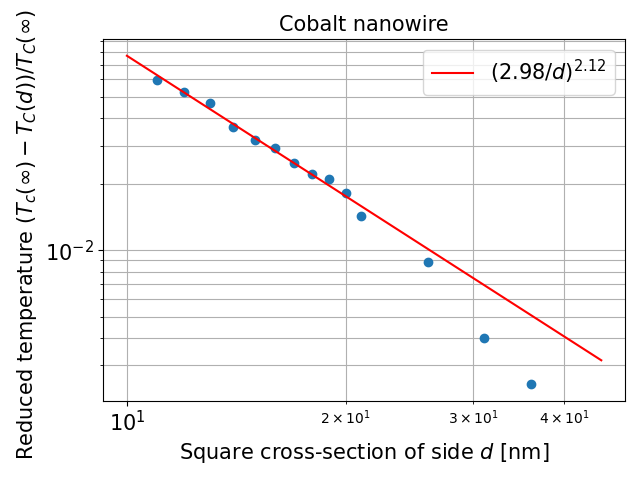}
\includegraphics[width=8cm]{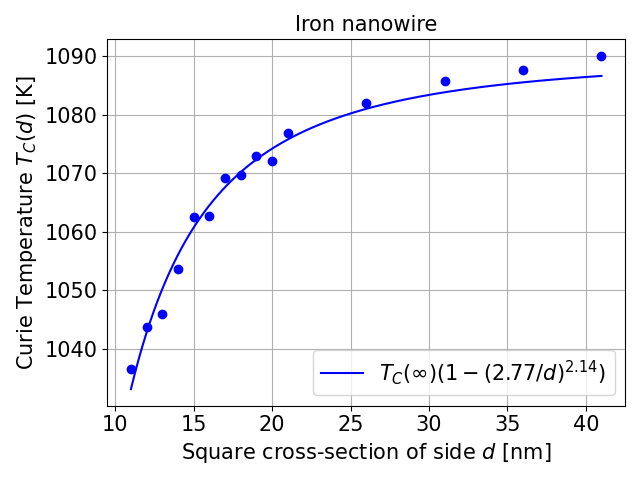}
\includegraphics[width=8cm]{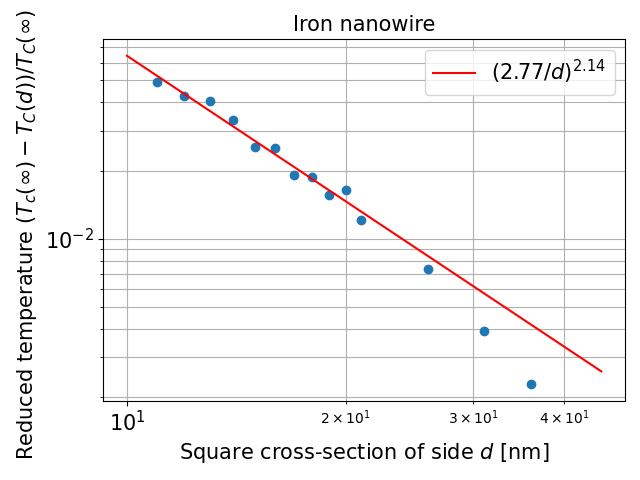}
\includegraphics[width=8cm]{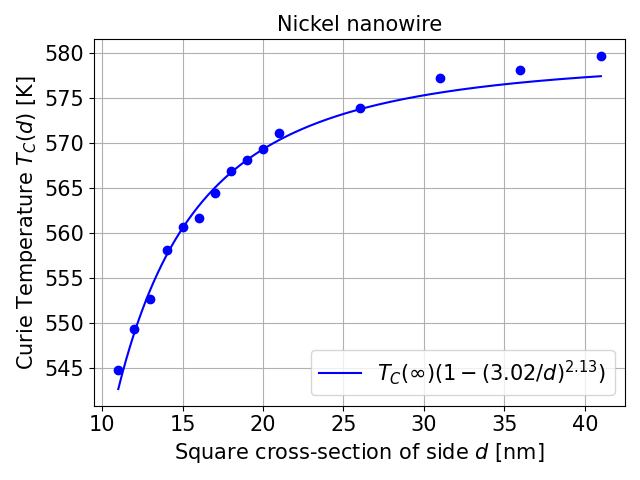}
\includegraphics[width=8cm]{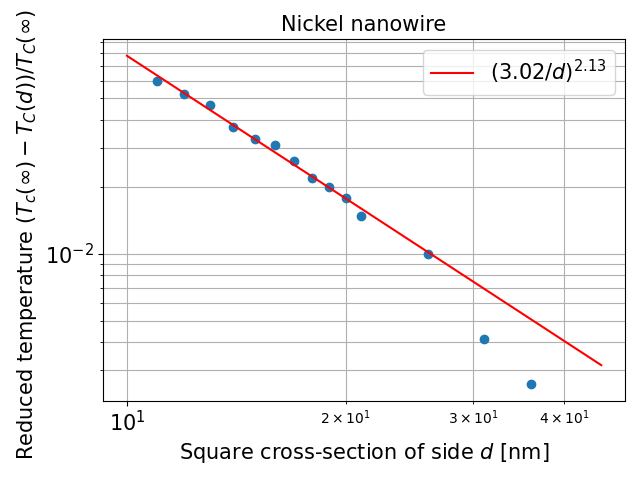}
\caption{Behavior of the Curie temperature as a function of the cross-section size $d$, for nanowires of different materials: cobalt (top panels), iron (middle panels) and nickel (bottom panels).  Numerical results are shown as blue dots, while the solid blue and red lines represent the theoretical power-law of Eq. \eqref{loi_T_C_en_fct_d_calcul_lambda}. The left column shows results on a linear scale, while the right column on a log-log scale.
The correlation length $\xi_0$ and critical exponent $\lambda$ can be read on each figure of the right column as: $(\xi_0/d)^{\lambda}.$ }
\label{fig_exponent_TC_nanowire}
\end{center}
\end{figure}

\begin{figure}[h!]
\begin{center}
\includegraphics[width=8cm]{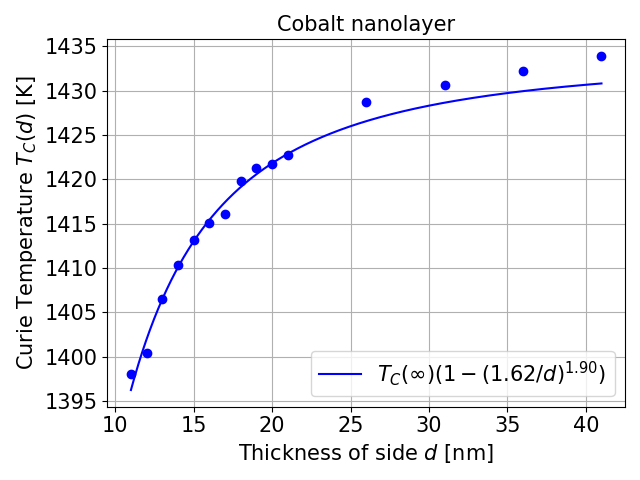}
\includegraphics[width=8cm]{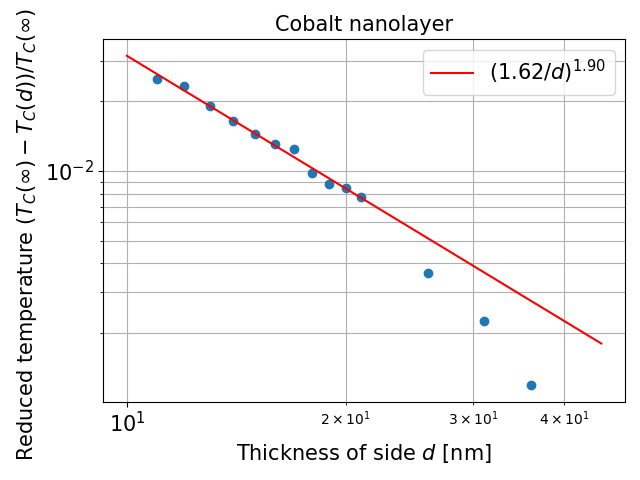}
\includegraphics[width=8cm]{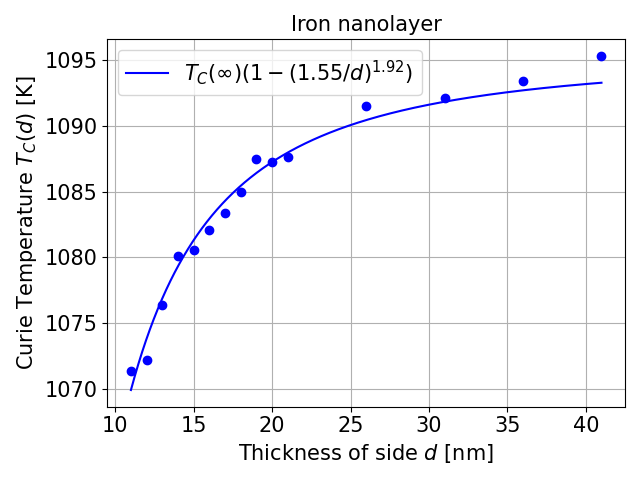}
\includegraphics[width=8cm]{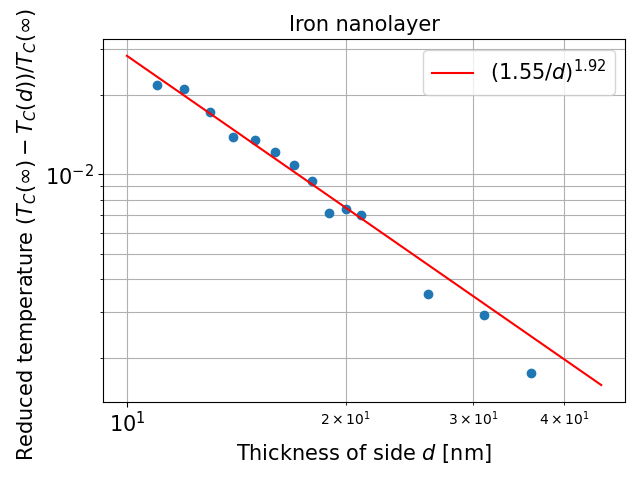}
\includegraphics[width=8cm]{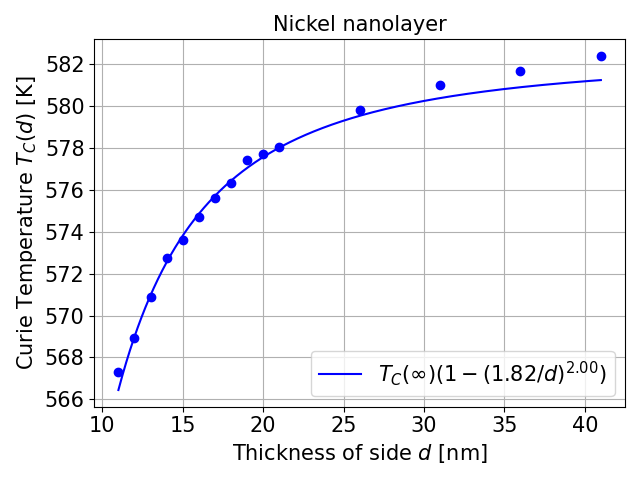}
\includegraphics[width=8cm]{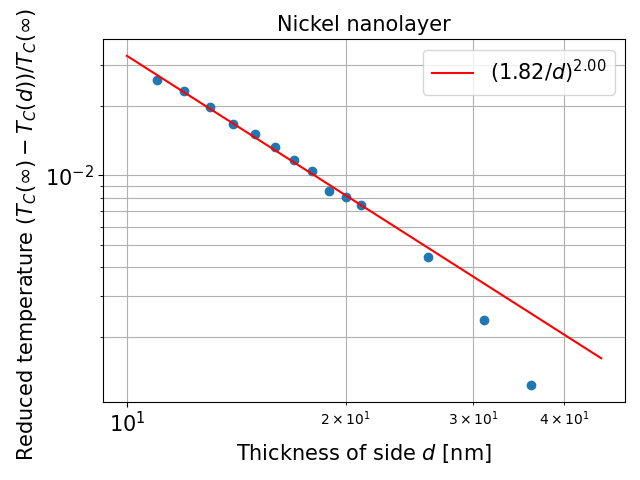}
\caption{Behavior of the Curie temperature as a function of the thickness $d$, for nanolayers of different materials: cobalt (top panels), iron (middle panels) and nickel (bottom panels). 
Numerical results are shown as blue dots, while the solid blue and red lines represent the theoretical power-law of Eq. \eqref{loi_T_C_en_fct_d_calcul_lambda}. The left column shows results on a linear scale, while the right column on a log-log scale.
The correlation length $\xi_0$ and critical exponent $\lambda$ can be read on each figure of the right column as: $(\xi_0/d)^{\lambda}.$}
\label{fig_exponent_TC_nanolayer}
\end{center}
\end{figure}

In the analysis of our simulation results, the Curie temperature of the bulk $T_{\text{C}}(\infty)$ is in fact replaced by the Curie temperature of the largest structure that we consider $T_{\text{C}}(d_{\text{max}})$, that is $d_{\text{max}}=41\,\rm nm $, see Table~\ref{table_temperature_curie}.
Figure~\ref{fig_exponent_TC_nanowire} (for nanowires) and Figure~\ref{fig_exponent_TC_nanolayer} (for nanolayers)
illustrate this power-law behaviour for the three materials considered here, both on a linear scale (left panels) and on a logarithmic scale (right panels). 
Blue circles correspond to the numerical results $T_{\text{C}}(d)$ extracted from Table~\ref{table_temperature_curie}. The red solid lines correspond to the theoretical power-law, Eq. \eqref{loi_T_C_en_fct_d_calcul_lambda}. The exponent $\lambda$ is deduced by fitting a log-log straight line through the numerical points using a least-square method, and then $\xi_0$ is obtained by finding the intercept $\lambda \log \xi_0$ of this line with the vertical axis.
The last two or three points further to the right deviate from the power-law, and were therefore discarded in the fitting procedure. 
The computed values of the correlation length $\xi_0$ and the critical exponent $\lambda$ are reported on each figure.

\GMcorr{The error estimate of the linear regression procedure is calculated with the standard error (SE) on the coefficients of the regression. Our regression curve is: $y_i=\lambda \log \xi_0 -\lambda x_i \equiv a+bx_i$, where $x_i = \log d_i$ and $y_i = \log\left(\frac{T_C(\infty)-T_C(d_i)}{T_C(\infty)}\right)$ for a given size $d_i$. We first compute the standard errors on $a$ and $b$ with the usual formulae:
\begin{equation*}
{\rm SE}(a) =\sqrt{\frac{\sum_{i=1}^{N_d}(y_i-\widehat{y}_i)^2}{(N_d-2)}\left(\frac{1}{N_d}+\frac{\bar{x}^2}{S_{xx}}\right)},
\end{equation*}
\begin{equation*}
{\rm SE}(b) =\sqrt{\frac{\sum_{i=1}^{N_d}(y_i-\widehat{y}_i)^2}{(N_d-2)S_{xx}}},
\label{eq_erreur_lambda}
\end{equation*}
where $S_{xx}=\sum_{i=1}^{N_d}(x_i-\bar{x})^2$, with $N_d$ the number of tested sizes $d$, and $\bar{x}=\frac{1}{N_d}\sum_{i=1}^{N_d}x_i$ denotes the mean of $x_i$, and $\widehat{y}_i$ is the estimate of $y_i$ obtained by the linear regression.
Then, the SEs on $\lambda$ and $\xi_0$ are computed as:
\begin{equation}
{\rm SE}(\lambda) = {\rm SE}(b) ; \quad{\rm SE}(\xi_0) =\frac{\xi_0}{\lambda} \left[{\rm SE}(a)+\log\xi_0\, {\rm SE}(b)\right].
\label{eq_erreur_xi0}
\end{equation}
These standard errors are listed in Table~\ref{tab:erreur_regression}.
}
\begin{table}[h]
\centering
\begin{tabular}{|lcc||lcc|}
\hline
\multicolumn{3}{|c||}{\textbf{Nanowire}}&\multicolumn{3}{c|}{\textbf{Nanolayer}}\\
Material& ${\rm SE(\lambda)}$ & ${\rm SE(\xi_0)}$ & Material& ${\rm SE(\lambda)}$ & ${\rm SE(\xi_0)}$ \\
\hline
Cobalt&$8.15\times10^{-2}$&$4.42\times10^{-1}$ &Cobalt&$6.37\times10^{-2}$&$1.76\times10^{-1}$\\
Iron&$1.17\times10^{-1}$&$5.75\times10^{-1}$&Iron&$1.03\times10^{-1}$&$2.68\times10^{-1}$\\
Nickel&$7.04\times10^{-2}$&$3.85\times10^{-1}$&Nickel&$4.98\times10^{-2}$&$1.52\times10^{-1}$\\
\hline
\end{tabular}
\caption{Standard errors (SE) on the coefficients $\lambda$ and $\xi_0$ of the linear regression, from Eq. \eqref{eq_erreur_xi0}.}
\label{tab:erreur_regression}
\end{table}
%
% \begin{table}[h]
%\centering
%\begin{tabular}{|lcc||lcc|}
%\hline
%\multicolumn{3}{|c||}{\textbf{Nanowire}}&\multicolumn{3}{c|}{\textbf{Nanolayer}}\\
%Material& Conf. int. for $\lambda$ & Conf. int. for $\xi_0$  & Material& Conf. int. for $\lambda$  & Conf. %int. for $\xi_0$  \\
%\hline
%Cobalt&$[1.94, 2.31]$&$[1.98, 3.98]$ &Cobalt&$[1.76, 2.04]$&$[1.22, 2.02]$\\
%Iron&$[1.87, 2.40]$&$[1.47, 4.07]$&Iron&$[1.68, 2.15]$&$[0.95, 2.16]$\\
%Nickel&$[1.97, 2.29]$&$[2.15, 3.89]$&Nickel&$[1.89, 2.12]$&$[1.48, 2.16]$\\
%\hline
%\end{tabular}
%\caption{95\% confidence intervals on the coefficients $\lambda$ and $\xi_0$, computed thanks to %Eq.~\eqref{eq_int_confiance}.}
%\label{tab:interval_confiance}
%\end{table}

In our simulations, the correlation length $\xi_0$ ranges from 2.77~nm to 3.02~nm for nanowires, while for nanolayers it varies from 1.55~nm to 1.82~nm. \GMcorr{The differences are mostly within the error range shown in Table 4}. These values are broadly in good agreement with those observed in the experiments.

As to the critical exponent, we obtain  $\lambda=2.12-2.14$ \GMcorr{(a range that is within the numerical uncertainty, see Table 4)} for all three materials in the case of nanowires. For nanolayers, we obtain $\lambda=1.90$, 1.92 and 2.00, for cobalt, iron and nickel, respectively. \GMcorr{Within the statistical errors (Table 4),} all exponents are compatible with the value $\lambda=2$, which is also the value observed in atomistic mean-field simulations \cite{Penny2019}.

\subsection{Nonequilibrium properties}

\begin{figure}[h!]
\begin{center}
\includegraphics[width=8cm]{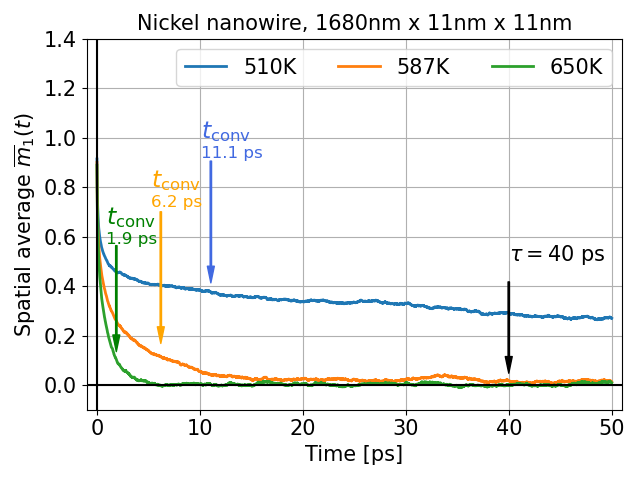}
\caption{Evolution of the spatially-averaged $x$ component of the magnetic moment $\overline{m}_1(t)$ for a nickel nanowire of cross-section $d=11$~nm, for three different temperatures, $T=510$~K, $T=587$~K and $T=650$~K. The corresponding convergence times ($t_{\text{conv}}=1.9$~ps, 6.2~ps, and 11.1~ps) are shown on the figure. We also show the transient time $\tau = 40$ ps, which is the same for all simulations (time used to average the total magnetization $M_{\text{tot}}$).
\label{fig_diff_tconv_tau}}
\end{center}
\end{figure}

Although we have used our code to study steady-state phenomena, such as the dependence of the magnetization curve and Curie temperature with size, our approach is fundamentally time-dependent. Indeed, to obtain our results we solved numerically the time-dependent LLG equation (with thermal effects) and then deduced steady-state quantities through time and/or ensemble averages. However, the code could be used to investigate nonequilibrium properties, such as transient behaviors, the motion of domain walls, instabilities, etc. 
Over more commonly used statistical methods, this dynamical approach has the further advantage of being well-suited to study the fluctuations that appear in the vicinity of phase transitions. These fluctuations play a key role in finite systems, particularly when their dimensionality is low.
This vast realm is left to future investigations. Here, we conclude our work by searching for a signature of the phase transition occurring at $T_{\text{C}}$ in the dynamical behaviour of the magnetization. We do this by looking at the convergence time towards the equilibrium state.

In Figure~\ref{fig_magnetization_en_fct_temps}, we observed that the convergence time to the plateau state depends on the temperature. 
More precisely, we define the {\em convergence time} as the first time at which the $x$ component of the spatially-averaged magnetic moment $\overline{m}_1(t)$ becomes sufficiently close to the total magnetization $M_{\text{tot}}$ (which corresponds to its plateau): 
\begin{equation}
t_{\text{conv}} = \underset{t\in[0, t_f]}{\text{inf}}  |\overline{m}_1(t)-M_{\text{tot}}|<\varepsilon,
\label{tconv}
\end{equation}
with the tolerance parameter $\varepsilon$ fixed to 0.1. Figure~\ref{fig_diff_tconv_tau} illustrates schematically this convergence time $t_{\text{conv}}$, and also shows the {\em transient time} $\tau$, which is the time used to compute the average magnetization, see Eq. \eqref{magnet_totale_2}.

\begin{figure}[h!]
\begin{center}
\includegraphics[width=8cm]{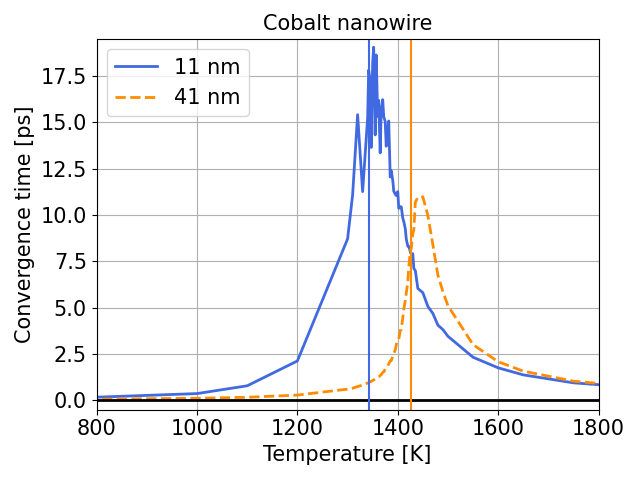}
\includegraphics[width=8cm]{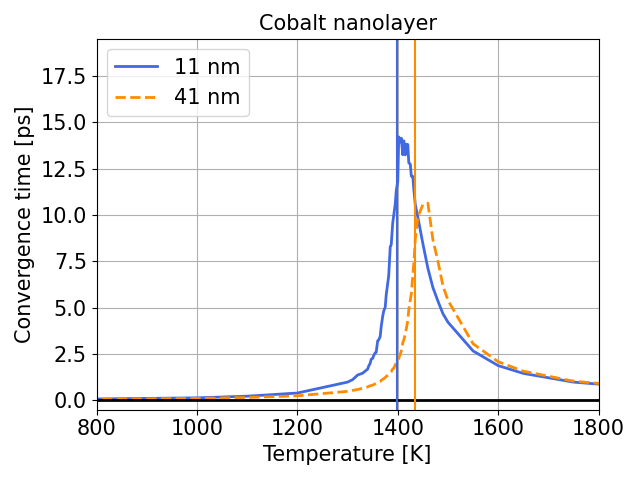}
\includegraphics[width=8cm]{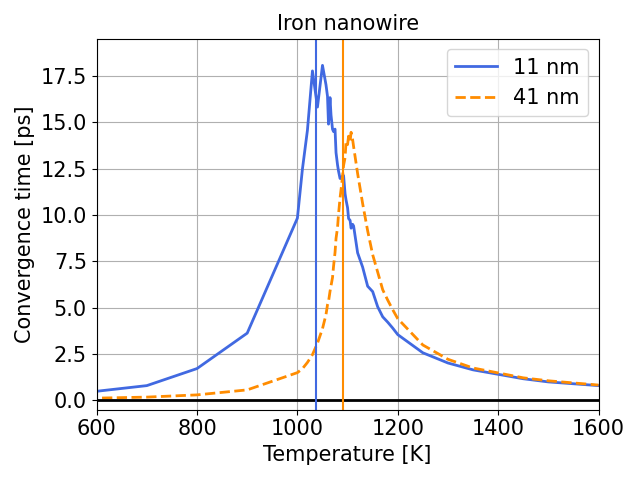}
\includegraphics[width=8cm]{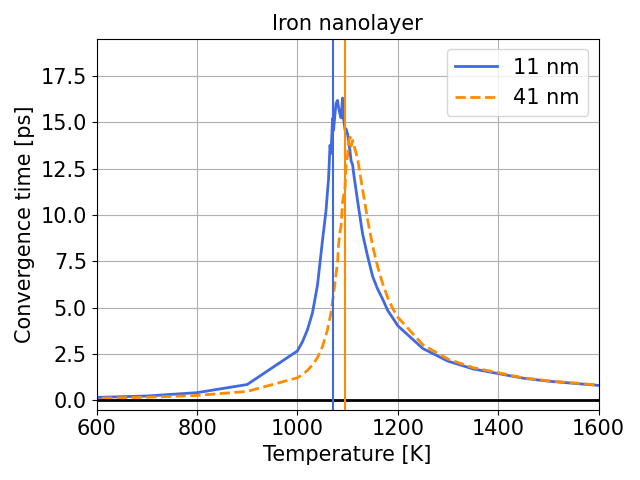}
\includegraphics[width=8cm]{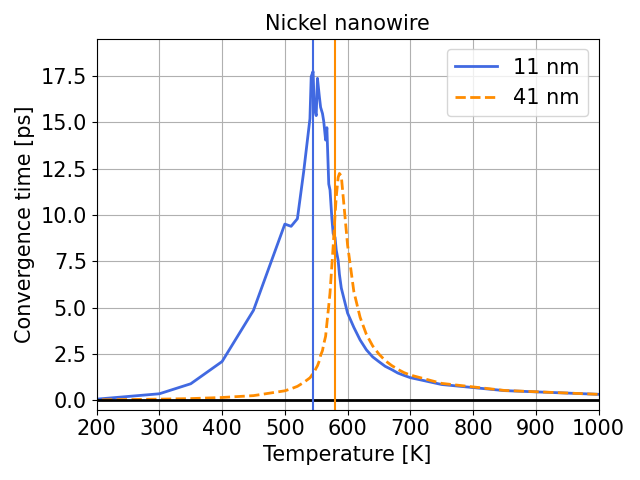}
\includegraphics[width=8cm]{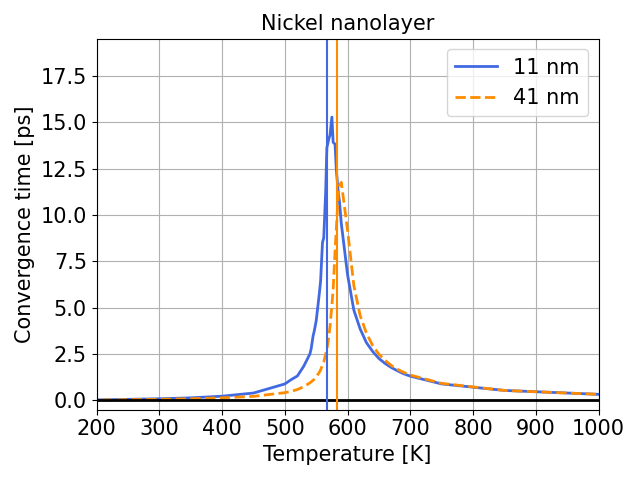}
\caption{Convergence time $t_{\text{conv}}$, from Eq. \eqref{tconv}, as a function of the temperature for cobalt (top panels), iron (middle panels) and nickel (bottom panels). Left column: nanowire geometry with $6000$ nm$\times d$ nm$\times d$ nm; right column: nanolayer geometry with $600$ nm$\times 600$ nm$\times d$ nm. Two values of $d$ are considered: $d=11$ nm (blue solid curve) and $d=41$ nm (orange dash curve). The vertical lines correspond to the numerical Curie temperatures reported in Table~\ref{table_temperature_curie}, for $d=11$ nm (blue vertical line) and $d=41$ nm (orange vertical line). \label{fig_transcient_time}}
\end{center}
\end{figure}

Figure~\ref{fig_transcient_time} shows the convergence time, as a function of the temperature, for  nanowires of size $6000\
\times\ d\ \times\ d$ nm$^3$ and nanolayers of size $600\ \times\ 600\ \times\ d$ nm$^3$, for two values of $d$.
First, we note that the convergence time $t_{\text{conv}}$ is always smaller than the transient time $\tau$ taken to compute the average $M_{\text{tot}}$ ($t_{\text{conv}}$ has a maximum value around 20 ps, which is always smaller than $\tau = 40$ ps). This confirms that convergence to the plateau state takes place before $\tau$ and hence that the calculation of $M_{\text{tot}}$ is correct. 

This figure also shows that the convergence time peaks near the Curie temperature \cite{Maurat2009}. This is a dynamical signature that large fluctuations near $T_{\text{C}}$ translate into longer times taken by the system to relax to its equilibrium state.
The rapid increase of the  relaxation time close to $T_{\text{C}}$ is known as critical
slowing down \cite{Peczak1993,Chubykalo2006}, an effect which is characteristic of second-order phase transitions. 

Moreover, Figure~\ref{fig_transcient_time} illustrates the effect of size on this phase transition. The structures with the smallest size ($d=11$~nm, blue curves) have longer convergence times $t_{\text{conv}}$ and larger widths than the larger structures ($d=41$~nm, orange curves). Also, the convergence times and the widths are larger for 1D nanowires than for 2D nanolayers.
One can deduce that fluctuations near the transition temperature $T_{\text{C}}$ are stronger for smaller and lower-dimensional structures.

\section{Conclusion}\label{Conclusion}

In this work, we developed a computational code (written in \textmd{Python}) that solves the LLG equation of micromagnetism, including finite-temperature effects. By adopting an appropriate temperature scaling with the computational cell size $\Delta x $ \cite{Hahn_2019}, it was possible to recover magnetization curves and Curie temperatures that match closely the experimental ones for cobalt, nickel and iron nano-objects. 
Compared to fully atomistic simulations, our micromagnetic approach has the advantage of being less computationally expensive, as it relies on a mesoscopic description at a scale $\ell$ much larger than the lattice constant: $\ell \gg \Delta x \gg a_{\text{eff}}$.
It is also more easily amenable to dynamical simulations to study, for instance, the motion of domain walls or other time-dependent phenomena.

Using this accurate computational tool, we investigated the size-dependence of the Curie temperature for 1D nanowires and 2D nanolayers, by varying the smallest size $d$ of the system. We confirmed that the difference between the computed finite-size $T_{\text{C}}$ and the bulk $T_{\text{C}}$ follows a power-law of the type: $(\xi_0/d)^\lambda$, where $\xi_0$ is the correlation length at zero temperature and $\lambda$ is a critical exponent. We obtained values of $\xi_0$ in the nanometer range, also in accordance with other simulations and experiments. The computed critical exponent was found to be close to $\lambda=2$ for all considered materials and geometries, which is the expected result for a mean-field approach, but slightly larger than the values observed experimentally. 
Finally, the time-dependent model developed here represents an effective tool for studying thermal fluctuations near the ferromagnetic phase transition.

All in all, the behaviour of 1D and 2D ferromagnetic nano-objects, as a function of both temperature and size, was recovered with good precision and a relatively low computational cost in comparison to fully atomistic simulations. This computational tool may therefore be applied in the future to more complex configurations, involving for instance 3D structures and dynamical effects.

\section*{Acknowledgements}
The authors gratefully acknowledge partial support by the ANR project MOSICOF ANR-21-CE40-0004. R. C\^ote acknowledges support from the University of Strasbourg Institute for Advanced Study (USIAS) for a Fellowship within the French national programme ``Investment for the future'' (IdEx-Unistra). 
This work of the Interdisciplinary Thematic Institute IRMIA++, as part of the ITI 2021-2028 program of the University of Strasbourg, CNRS and Inserm, was supported by IdEx Unistra (ANR-10-IDEX-0002), and by SFRI-STRAT’US project (ANR-20-SFRI-0012) under the framework of the French Investments for the Future Program.
The authors would like to thank J\'er\^ome Lelong and Bertrand Dup\'e for a suggestion that helped improve the computer code, and Guillaume Ferriere, Ludovic Godard-Cadillac and Yannick Privat for the enriching discussions  throughout the project.
We also thank Riccardo Hertel for his thorough reading of the manuscript and many insightful comments.

%%%%%%%%%%%%%%%%%%%%%%%%%%%%%%
\bibliographystyle{abbrv}
\bibliography{Biblio.bib}

\newpage

\begin{center}
\textsc{cl\'ementine court\`es}\\
Universit\'e de Strasbourg \& Inria Nancy Grand Est\\
CNRS, IRMA UMR 7501,\\
7 rue Ren\'e Descartes,\\
67084 Strasbourg, France\\
{\tt clementine.courtes@unistra.fr}\\
\vspace*{0.5cm}

\textsc{matthieu boileau}\\
CNRS \& Inria Nancy Grand Est\\
CNRS, IRMA UMR 7501,\\
67084 Strasbourg, France\\
{\tt matthieu.boileau@math.unistra.fr}\\
\vspace*{0.5cm}

\textsc{rapha\"el c\^ote}\\
Universit\'e de Strasbourg \& University of Strasbourg Institute for Advanced Study\\
CNRS, IRMA UMR 7501,\\
7 rue Ren\'e Descartes,\\
67084 Strasbourg, France\\
{\tt raphael.cote@unistra.fr}\\
\vspace*{0.5cm}

\textsc{paul-antoine hervieux}\\
Universit\'e de Strasbourg \\
CNRS, Institut de Physique et Chimie des Matériaux de Strasbourg, UMR 7504\\
F-67000 Strasbourg, France\\
{\tt paul-antoine.hervieux@ipcms.unistra.fr}\\
\vspace*{0.5cm}

\textsc{giovanni manfredi}\\
Universit\'e de Strasbourg \\
CNRS, Institut de Physique et Chimie des Matériaux de Strasbourg, UMR 7504\\
F-67000 Strasbourg, France\\
{\tt giovanni.manfredi@ipcms.unistra.fr}\\
\end{center}

\end{document}